# Capacity-Achieving Ensembles for the Binary Erasure Channel With Bounded Complexity*

Henry D. Pfister
Qualcomm, Inc.
CA 92121, USA
`hpfister@qualcomm.com`

Igal Sason
Technion
Haifa 32000, Israel
`Sason@ee.technion.ac.il`

Rüdiger Urbanke
EPFL
Lausanne 1015, Switzerland
`Rudiger.Urbanke@epfl.ch`

September 8, 2004

**Abstract**

We present two sequences of ensembles of non-systematic irregular repeat-accumulate codes which asymptotically (as their block length tends to infinity) achieve capacity on the binary erasure channel (BEC) with *bounded complexity* per information bit. This is in contrast to all previous constructions of capacity-achieving sequences of ensembles whose complexity grows at least like the log of the inverse of the gap (in rate) to capacity. The new bounded complexity result is achieved by puncturing bits, and allowing in this way a sufficient number of state nodes in the Tanner graph representing the codes. We also derive an information-theoretic lower bound on the decoding complexity of randomly punctured codes on graphs. The bound holds for every memoryless binary-input output-symmetric channel and is refined for the BEC.

## 1 Introduction

During the last decade, there have been many exciting developments in the construction of low-complexity error-correction codes which closely approach the capacity of many standard communication channels with feasible complexity. These codes are understood to be codes defined on graphs, together with the associated iterative decoding algorithms. By now, there is a large collection of these codes that approach the channel capacity quite closely with moderate complexity.

The first capacity-achieving sequences of ensembles of low-density parity-check (LDPC) codes for the binary erasure channel (BEC) were found by Luby et al. [7, 8] and Shokrollahi [16]. Following these pioneering works, Oswald and Shokrollahi presented in [9] a systematic study of capacity-achieving degree distributions (d.d.) for sequences of ensembles of LDPC codes whose transmission takes place over the BEC. Capacity-achieving ensembles of irregular repeat-accumulate (IRA) codes

---

for the BEC were introduced and analyzed in [4, 15], and also capacity-achieving ensembles for erasure channels with memory were designed and analyzed in [10, 11].

In [5], Khandekar and McEliece discussed the complexity of achieving the channel capacity on the BEC, and more general channels with vanishing bit error probability. They conjectured that if the achievable rate under message-passing iterative (MPI) decoding is a fraction $1-\varepsilon$ of the channel capacity, then for a wide class of channels, the encoding complexity scales like $\ln \frac{1}{\varepsilon}$ and the decoding complexity scales like $\frac{1}{\varepsilon} \ln \frac{1}{\varepsilon}$. This conjecture is based on the assumption that the number of edges (per information bit) in the associated bipartite graph scales like $\ln \frac{1}{\varepsilon}$, and the required number of iterations under MPI decoding scales like $\frac{1}{\varepsilon}$. However, for codes defined on graphs which are transmitted over a BEC, the decoding complexity under the MPI algorithm behaves like $\ln \frac{1}{\varepsilon}$ (same as encoding complexity) [7, 14, 16]. This is since the absolute reliability provided by the BEC allows every edge in the graph to be used only once during MPI decoding.

In [14], Sason and Urbanke considered the question of how sparse can parity-check matrices of binary linear codes be, as a function of their gap (in rate) to capacity (where this gap depends on the channel and the decoding algorithm). If the code is represented by a standard Tanner graph without state nodes, the decoding complexity under MPI decoding is strongly linked to the density of the corresponding parity-check matrix (i.e., the number of edges in the graph per information bit). In particular, they considered an arbitrary sequence of binary linear codes which achieves a fraction $1-\varepsilon$ of the capacity of a memoryless binary-input output-symmetric (MBIOS) channel with vanishing bit error probability. By information-theoretic tools, they proved that for every such sequence of codes and every sequence of parity-check matrices which represent these codes, the asymptotic density of the parity-check matrices grows at least like $\frac{K_1 + K_2 \ln \frac{1}{\varepsilon}}{1-\varepsilon}$ where $K_1$ and $K_2$ are constants which were given explicitly as a function of the channel statistics (see [14, Theorem 2.1]). It is important to mention that this bound is valid under ML decoding, and hence, it also holds for every sub-optimal decoding algorithm. The tightness of the lower bound for MPI decoding on the BEC was demonstrated in [14, Theorem 2.3] by analyzing the capacity-achieving sequence of check-regular LDPC-code ensembles introduced by Shokrollahi [16]. Based on the discussion in [14], it follows that for every iterative decoder which is based on the standard Tanner graph, there exists a fundamental tradeoff between performance and complexity, and the complexity (per information bit) becomes *unbounded* when the gap between the achievable rate and the channel capacity vanishes. Therefore, it was suggested in [14] to study if better performance versus complexity tradeoffs can be achieved by allowing more complicated graphical models (e.g., graphs which also involve state nodes).

In this paper, we present sequences of capacity-achieving ensembles for the BEC with bounded complexity under MPI decoding. The new ensembles are non-systematic IRA codes with properly chosen d.d. (for background on IRA codes, see [4] and Section 2). The new bounded complexity results improve on the results in [15], and demonstrate the superiority of properly designed non-systematic IRA codes over systematic IRA codes (since with probability 1, the complexity of any sequence of ensembles of systematic IRA codes becomes *unbounded* under MPI decoding when the gap between the achievable rate and the capacity vanishes [15, Theorem 1]). The new bounded complexity result is achieved by allowing a sufficient number of state nodes in the Tanner graph representing the codes. Hence, it answers in the affirmative a fundamental question which was posed in [14] regarding the impact of state nodes in the graph on the performance versus complexity tradeoff under MPI decoding. We suggest a particular sequence of capacity-achieving ensembles of non-systematic IRA codes where the degree of the parity-check nodes is 5, so the complexity per information bit under MPI decoding is equal to $\frac{5}{1-p}$ when the gap (in rate) to capacity vanishes ($p$ designates the bit erasure probability of the BEC). Computer simulation results for these ensembles appear to agree with this analytical result. It is worth noting that our method of truncating the check d.d. is similar to the bi-regular check d.d. introduced in [18] for non-systematic IRA codes.

We also derive in this paper an information-theoretic lower bound on the decoding complexity of randomly punctured codes on graphs. The bound holds for every MBIOS channel with a refinement for the particular case of a BEC.

The structure of the paper is as follows: Section 2 provides preliminary material on ensembles of IRA codes, Section 3 presents our main results which are proved in Section 4. Analytical and numerical results for the considered degree distributions and their asymptotic behavior are discussed in Section 5. Practical considerations and simulation results for our ensembles of IRA codes are presented in Section 6. We conclude our discussion in Section 7. Three appendices also present important mathematical details which are related to Sections 4 and 5.

## 2   IRA Codes

We consider here ensembles of non-systematic IRA codes. We assume that all information bits are punctured. The Tanner graph of these codes is shown in Fig. 1. These codes can be viewed as serially concatenated codes, where the outer code is a mixture of repetition codes of varying order and the inner code is generated by a differential encoder with puncturing. We define these ensembles by a uniform choice of the interleaver separating the component codes.

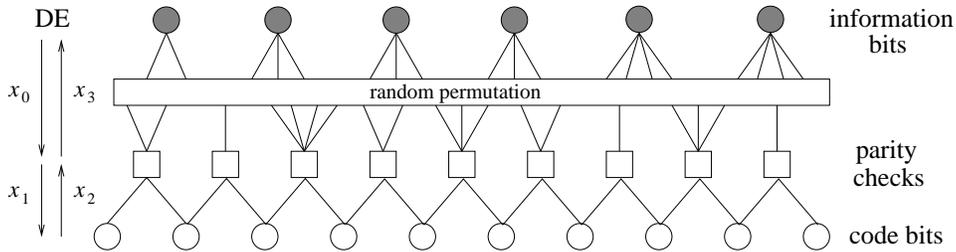

Figure 1: The Tanner graph of IRA codes.

Using standard notation, an ensemble of IRA codes is characterized by its block length $n$ and its d.d. pair $\lambda(x) = \sum_{i=1}^{\infty} \lambda_i x^{i-1}$ and $\rho(x) = \sum_{i=1}^{\infty} \rho_i x^{i-1}$. Here, $\lambda_i$ (or $\rho_i$, respectively) designates the probability that a randomly chosen edge (among the edges that connect the information nodes and the parity-check nodes) is connected to an information bit node (or to a parity-check node) of degree $i$. As is shown in Fig. 1, every parity-check node is also connected to two code bits; this is a result of the differential encoder which is the inner code of these serially concatenated and interleaved codes. Let $R(x) = \sum_{i=1}^{\infty} R_i \, x^i$ be a power series where the coefficient $R_i$ denotes the fraction of parity-check nodes that are connected to $i$ information nodes. Therefore, $R(\cdot)$ and $\rho(\cdot)$ are related by the equation

$$R(x) = \frac{\int_0^x \rho(t) \, \mathrm{d}t}{\int_0^1 \rho(t) \, \mathrm{d}t}. \qquad (1)$$

We assume that the random permutation in Fig. 1 is chosen with equal probability from the set of all permutations. A randomly selected code from this ensemble is used to communicate over a BEC with erasure probability $p$. The asymptotic performance of the MPI decoder (as the block length tends to infinity) can be analyzed by tracking the average fraction of erasure messages which are passed in the graph of Fig. 1 during the $l^{\text{th}}$ iteration. This technique was introduced in [12] and is known as density evolution (DE). In the asymptotic case where the block length tends to infinity, the messages which are passed through the edges of the Tanner graph are statistically independent.

Using the same notation as in [4], let $x_0^{(l)}$ be the probability of erasure for a message from information nodes to parity-check nodes, $x_1^{(l)}$ be the probability of erasure from parity-check nodes to code nodes, $x_2^{(l)}$ be the probability of erasure from code nodes to parity-check nodes, and finally, let $x_3^{(l)}$ be the probability of erasure for messages from parity-check nodes to information nodes (see Fig. 1). We now assume that we are at a fixed point of the MPI decoding algorithm, and solve for $x_0$. We obtain the following equations:

$$x_1 = 1 - (1 - x_2)R(1 - x_0), \tag{2}$$
$$x_2 = px_1, \tag{3}$$
$$x_3 = 1 - (1 - x_2)^2 \rho(1 - x_0), \tag{4}$$
$$x_0 = \lambda(x_3). \tag{5}$$

The only difference between equations (2)–(5) and those in [4] is the absence of a $p$ in (5). This modification stems from the fact that all information bits are punctured in the ensemble considered. Solving this set of equations for a fixed point of iterative decoding provides the equation

$$x_0 = \lambda \left( 1 - \left[ \frac{1-p}{1 - pR(1-x_0)} \right]^2 \rho(1 - x_0) \right). \tag{6}$$

If Eq. (6) has no solution in the interval $(0, 1]$, then according to the DE analysis of MPI decoding, the bit erasure probability must converge to zero. Therefore, the condition that

$$\lambda \left( 1 - \left[ \frac{1-p}{1 - pR(1-x)} \right]^2 \rho(1 - x) \right) < x, \quad \forall x \in (0, 1], \tag{7}$$

implies that MPI decoding obtains a vanishing bit erasure probability as the block length tends to infinity.

The design rate of the ensemble of non-systematic IRA codes can be computed by matching edges in the Tanner graph shown in Fig. 1. In particular, the number of edges in the permutation must be equal to both the number of information bits times the average information bit degree and the number of code bits times the average parity-check degree (ignoring the parity-check edges connected to the differential encoder). This implies that the design rate is equal to

$$R^{\mathrm{IRA}} = \frac{\int_0^1 \lambda(x)\, dx}{\int_0^1 \rho(x)\, dx}. \tag{8}$$

Furthermore, we will see that $R^{\mathrm{IRA}} = 1 - p$ for any pair of d.d. $(\lambda, \rho)$ which satisfies Eq. (6) for all $x_0 \in [0, 1]$.

In order to find a capacity-achieving ensemble of IRA codes, we generally start by finding a d.d. pair $(\lambda, \rho)$ with non-negative power series expansions which satisfies Eq. (6) for all $x_0 \in [0, 1]$. Next, we slightly modify $\lambda(\cdot)$ or $\rho(\cdot)$ so that Eq. (7) is satisfied and the new design rate in Eq. (8) is equal to $(1 - \varepsilon)(1 - p)$ for an arbitrarily small $\varepsilon > 0$. Since the capacity of the BEC is $1 - p$, this gives an ensemble which has vanishing bit erasure probability under MPI decoding at rates which are arbitrarily close to capacity.

# 3 Main Results

**Definition 1.** Let $\{\mathcal{C}_m\}$ be a sequence of binary linear codes of rate $R_m$, and assume that for every $m$, the codewords of the code $\mathcal{C}_m$ are transmitted with equal probability over a channel whose capacity is $C$. This sequence is said to *achieve a fraction $1 - \varepsilon$ of the channel capacity with vanishing bit error probability* if $\lim_{m \to \infty} R_m \geq (1 - \varepsilon)C$, and there exists a decoding algorithm under which the average bit error probability of the code $\mathcal{C}_m$ tends to zero in the limit where $m$ tends to infinity.[1]

**Definition 2.** Let $\mathcal{C}$ be an ensemble of LDPC or IRA codes whose d.d., $\lambda(\cdot)$ and $\rho(\cdot)$, can be chosen arbitrarily (subject to possibly some constraints). The *encoding and the decoding complexity* are measured in operations per information bit which are required for achieving a fraction $1 - \varepsilon$ of the channel capacity with vanishing bit erasure probability. Unless the pair of d.d. is specified, the encoding and the decoding complexity are measured with respect to *the best ensemble* (i.e., for the optimized pair of d.d.), and refer to the average complexity over this ensemble (as the block length of the codes tends to infinity, the complexity of a typical code from this ensemble concentrates to the average complexity). We denote the encoding and the decoding complexity by $\chi_E(\varepsilon, \mathcal{C})$ and $\chi_D(\varepsilon, \mathcal{C})$, respectively.

We note that for the BEC, both the encoding and the decoding complexity of IRA codes under MPI decoding is equal to the normalized number of edges per information bit in the associated Tanner graph.

**Theorem 1 (Bit-Regular Ensembles with Bounded Complexity).** Consider the ensemble of bit-regular non-systematic IRA codes $\mathcal{C}$, where the d.d. of the information bits is given by

$$\lambda(x) = x^{q-1}, \quad q \geq 3 \tag{9}$$

which implies that each information bit is repeated $q$ times. Assume that the transmission takes place over a BEC with erasure probability $p$, and let the d.d. of the parity-check nodes[2] be

$$\rho(x) = \frac{1 - (1-x)^{\frac{1}{q-1}}}{\left[1 - p\left(1 - qx + (q-1)\left[1 - (1-x)^{\frac{q}{q-1}}\right]\right)\right]^2}. \tag{10}$$

Let $\rho_n$ be the coefficient of $x^{n-1}$ in the power series expansion of $\rho(x)$ and, for an arbitrary $\epsilon \in (0,1)$, define $M(\varepsilon)$ to be the smallest positive integer[3] $M$ such that

$$\sum_{n=M+1}^{\infty} \rho_n < \frac{\varepsilon}{q(1-p)}. \tag{11}$$

The $\varepsilon$-truncated d.d. of the parity-check nodes is given by

$$\rho_\varepsilon(x) = \left(1 - \sum_{n=2}^{M(\varepsilon)} \rho_n\right) + \sum_{n=2}^{M(\varepsilon)} \rho_n x^{n-1}. \tag{12}$$

---

[1] We refer to vanishing bit erasure probability for the particular case of a BEC.

[2] The d.d. of the parity-check nodes refers only to the connection of the parity-check nodes with the information nodes. Every parity-check node is also connected to *two code bits* (see Fig. 1), but this is not included in $\rho(x)$.

[3] The existence of $M(\varepsilon)$ for $\varepsilon \in (0,1)$ follows from the fact that $\rho_n = O(n^{-q/(q-1)})$. This implies that $\sum_{n=M+1}^{\infty} \rho_n$ can be made arbitrarily small by increasing $M$.

For $q = 3$ and $p \in (0, \frac{1}{13}]$, the polynomial $\rho_\varepsilon(\cdot)$ has only non-negative coefficients, and the d.d. pair $(\lambda, \rho_\varepsilon)$ achieves a fraction $1-\varepsilon$ of the channel capacity with vanishing bit erasure probability under MPI decoding. Moreover, the complexity (per information bit) of encoding and decoding satisfies

$$\chi_E(\varepsilon, \mathcal{C}) = \chi_D(\varepsilon, \mathcal{C}) < q + \frac{2}{(1-p)(1-\varepsilon)}. \tag{13}$$

In the limit where $\varepsilon$ tends to zero, the capacity is achieved with a *bounded* complexity of $q + \frac{2}{1-p}$.

**Theorem 2 (Check-Regular Ensemble with Bounded Complexity).** Consider the ensemble of check-regular non-systematic IRA codes $\mathcal{C}$, where the d.d. of the parity-check nodes is given by

$$\rho(x) = x^2. \tag{14}$$

Assume that the transmission takes place over a BEC with erasure probability $p$, and let the d.d. of the information bit nodes be[4]

$$\lambda(x) = 1 + \frac{2p(1-x)^2 \sin\left(\frac{1}{3} \arcsin\left(\sqrt{-\frac{27p(1-x)^{\frac{3}{2}}}{4(1-p)^3}}\right)\right)}{\sqrt{3}\,(1-p)^4 \left(-\frac{p(1-x)^{\frac{3}{2}}}{(1-p)^3}\right)^{\frac{3}{2}}}. \tag{15}$$

Let $\lambda_n$ be the coefficient of $x^{n-1}$ in the power series expansion of $\lambda(x)$ and, for an arbitrary $\epsilon \in (0,1)$, define $M(\varepsilon)$ to be the smallest positive integer[5] $M$ such that

$$\sum_{n=M+1}^{\infty} \frac{\lambda_n}{n} < \frac{(1-p)\varepsilon}{3}. \tag{16}$$

This infinite bit d.d. is truncated by treating all information bits with degree greater than $M(\varepsilon)$ as pilot bits (i.e., these information bits are set to zero). Let $\lambda_\varepsilon(x)$ be the $\varepsilon$-truncated d.d. of the bit nodes. Then, for all $p \in [0, 0.95]$, the polynomial $\lambda_\varepsilon(\cdot)$ has only non-negative coefficients, and the modified d.d. pair $(\lambda_\varepsilon, \rho)$ achieves a fraction $1-\varepsilon$ of the channel capacity with vanishing bit erasure probability under MPI decoding. Moreover, the complexity (per information bit) of encoding and decoding is *bounded* and satisfies

$$\chi_E(\varepsilon, \mathcal{C}) = \chi_D(\varepsilon, \mathcal{C}) < \frac{5}{(1-p)(1-\varepsilon)}. \tag{17}$$

In the limit as $\varepsilon$ tends to zero, the capacity is achieved with a *bounded* complexity of $\frac{5}{1-p}$.

The following two conjectures extend Theorems 1 and 2 to a wider range of parameters. Both of these conjectures can be proved by showing that the power series expansions of $\lambda(x)$ and $\rho(x)$ are non-negative for this wider range. Currently, we can show that the power series expansions of $\lambda(x)$ and $\rho(x)$ are non-negative over this wider range only for small values of $n$ (using numerical methods) and large values of $n$ (using asymptotic expansions). We note that if these conjectures hold, then Theorem 1 is extended to the range $p \in [0, \frac{3}{13}]$ (as $q \to \infty$), and Theorem 2 is extended to the entire range $p \in [0, 1)$.

---
[4]For real numbers, one can simplify the expression of $\lambda(x)$ in (15). However, since we consider later $\lambda(\cdot)$ as a function of a complex argument, we prefer to leave it in the form of (15).

[5]The existence of $M(\varepsilon)$ for $\varepsilon \in (0,1)$ follows from the fact that $\lambda_n = O(n^{-3/2})$. This implies that $\sum_{n=M+1}^{\infty} \lambda_n/n$ can be made arbitrarily small by increasing $M$.

**Conjecture 1.** The result of Theorem 1 also holds for $q \geq 4$ if

$$p \leq \begin{cases} \dfrac{6 - 7q + 2q^2}{6 - 13q + 8q^2} & 4 \leq q \leq 8 \\ \dfrac{12 - 17q + 6q^2}{12 - 37q + 26q^2} & q \geq 9 \end{cases}. \tag{18}$$

We note that the form of Eq. (18) is implied by the analysis in Appendix A.

**Conjecture 2.** The result of Theorem 2 also holds for $p \in (0.95, 1)$.

In continuation to Theorem 2 and Conjecture 2, it is worth noting that Appendix C suggests a conceptual proof which in general could enable one to verify the non-negativity of the d.d. coefficients $\{\lambda_n\}$ for $p \in [0, 1 - \varepsilon]$, where $\varepsilon > 0$ can be made arbitrarily small. This proof requires though to verify the positivity of a fixed number of the d.d. coefficients, where this number grows considerably as $\varepsilon$ tends to zero. We chose to verify it for all $n \in \mathbb{N}$ and $p \in [0, 0.95]$. We note that a direct numerical calculation of $\{\lambda_n\}$ for small to moderate values of $n$, and the asymptotic behavior of $\lambda_n$ (which is derived in Appendix B) strongly supports Conjecture 2.

**Theorem 3 (Information-Theoretic Bound on the Complexity of Punctured Codes over the BEC).** Let $\{\mathcal{C}'_m\}$ be a sequence of binary linear block codes, and let $\{\mathcal{C}_m\}$ be a sequence of codes which is constructed by randomly puncturing information bits from the codes in $\{\mathcal{C}'_m\}$.[6] Let $P_{\text{pct}}$ designate the puncturing rate of the information bits, and suppose that the communication of the punctured codes takes place over a BEC with erasure probability $p$, and that the sequence $\{\mathcal{C}_m\}$ achieves a fraction $1 - \varepsilon$ of the channel capacity with vanishing bit erasure probability. Then with probability 1 w.r.t. the random puncturing patterns, and for an arbitrary representation of the sequence of codes $\{\mathcal{C}'_m\}$ by Tanner graphs, the asymptotic decoding complexity under MPI decoding satisfies

$$\liminf_{m \to \infty} \chi_D(\mathcal{C}_m) \geq \frac{p}{1-p} \left( \frac{\ln\left(\frac{P_{\text{eff}}}{\varepsilon}\right)}{\ln\left(\frac{1}{1-P_{\text{eff}}}\right)} + l_{\min} \right) \tag{19}$$

where

$$P_{\text{eff}} \triangleq 1 - (1 - P_{\text{pct}})(1 - p) \tag{20}$$

and $l_{\min}$ designates the minimum number of edges which connect a parity-check node with the nodes of the parity bits.[7] Hence, a necessary condition for a sequence of randomly punctured codes $\{\mathcal{C}_m\}$ to achieve the capacity of the BEC with *bounded complexity* is that the puncturing rate of the information bits satisfies the condition $P_{\text{pct}} = 1 - O(\varepsilon)$.

Theorem 4 suggests an extension of Theorem 3, though as will be clarified later, the lower bound in Theorem 3 is at least twice larger than the lower bound in Theorem 4 when applied to the BEC.

---

[6] Since we do not require that the sequence of original codes $\{\mathcal{C}'_m\}$ is represented in a systematic form, then by saying 'information bits', we just refer to any set of bits in the code $\mathcal{C}'_m$ whose size is equal to the dimension of the code and whose corresponding columns in the parity-check matrix are linearly independent. If the sequence of the original codes $\{\mathcal{C}'_m\}$ is systematic (e.g., turbo or IRA codes before puncturing), then it is natural to define the information bits as the systematic bits of the code.
[7] The fact that the value of $l_{\min}$ can be changed according to the choice of the information bits is a consequence of the bounding technique.

**Theorem 4 (Information-Theoretic Bound on the Complexity of Punctured Codes: General Case).** Let $\{\mathcal{C}'_m\}$ be a sequence of binary linear block codes, and let $\{\mathcal{C}_m\}$ be a sequence of codes which is constructed by randomly puncturing information bits from the codes in $\{\mathcal{C}'_m\}$. Let $P_{\mathrm{pct}}$ designate the puncturing rate of the information bits, and suppose that the communication takes place over an MBIOS channel whose capacity is equal to $C$ bits per channel use. Assume that the sequence of punctured codes $\{\mathcal{C}_m\}$ achieves a fraction $1-\varepsilon$ of the channel capacity with vanishing bit error probability. Then with probability 1 w.r.t. the random puncturing patterns, and for an arbitrary representation of the sequence of codes $\{\mathcal{C}'_m\}$ by Tanner graphs, the asymptotic decoding complexity per iteration under MPI decoding satisfies

$$\liminf_{m\to\infty} \chi_D(\mathcal{C}_m) \geq \frac{1-C}{2C} \frac{\ln\left(\frac{1}{\varepsilon} \frac{1-(1-P_{\mathrm{pct}})C}{2C \ln 2}\right)}{\ln\left(\frac{1}{(1-P_{\mathrm{pct}})(1-2w)}\right)} \tag{21}$$

where

$$w \triangleq \frac{1}{2} \int_{-\infty}^{+\infty} \min\left(f(y), f(-y)\right) \, \mathrm{d}y \tag{22}$$

and $f(y) \triangleq p(y|x=1)$ designates the conditional *pdf* of the channel, given the input is $x=1$. Hence, a necessary condition for a sequence of randomly punctured codes $\{\mathcal{C}_m\}$ to achieve the capacity of an MBIOS channel with *bounded complexity per iteration* under MPI decoding is that the puncturing rate of the information bits satisfies $P_{\mathrm{pct}} = 1 - O(\varepsilon)$.

**Remark 1 (Deterministic Puncturing).** It is worth noting that Theorems 3 and 4 both depend on the assumption that the set of information bits to be punctured is chosen randomly. It is an interesting open problem to derive information-theoretic bounds that apply to *every puncturing pattern* (including the best carefully designed puncturing pattern for a particular code). We also note that for any deterministic puncturing pattern which causes each parity-check to involve at least one punctured bit, the bounding technique which is used in the proofs of Theorems 3 and 4 becomes trivial and does not provide a meaningful lower bound on the complexity in terms of the gap (in rate) to capacity.

# 4 Proof of the Main Theorems

In this section, we prove our main theorems. The first two theorems are similar and both prove that under MPI decoding, specific sequences of ensembles of non-systematic IRA codes achieve the capacity of the BEC with bounded complexity (per information bit). The last two theorems provide an information-theoretic lower bound on the decoding complexity of randomly punctured codes on graphs. The bound holds for every MBIOS channel and is refined for the particular case of a BEC.

The approach used in the first two theorems was pioneered in [7] and can be broken into roughly three steps. The first step is to find a (possibly parameterized) d.d. pair $(\lambda, \rho)$ which satisfies the DE equation (6). The second step involves constructing an infinite set of parameterized (e.g., truncated or perturbed) d.d. pairs which satisfy the DE inequality (7). The third step is to verify that all of coefficients of the d.d. pair $(\lambda, \rho)$ are non-negative and sum to one for the parameter values of interest. Finally, if the design rate of the ensemble approaches $1-p$ for some limit point of the parameter set, then the ensemble achieves the channel capacity with vanishing bit erasure probability. The following lemma simplifies the proof of Theorems 1 and 2. We note that its proof is based on the analysis of capacity-achieving sequences for the BEC in [16] and the extension to erasure channels with memory in [10, 11].

**Lemma 1.** Any pair of d.d. functions $(\lambda, \rho)$ which satisfy $\lambda(0) = 0$, $\lambda(1) = 1$, and satisfy the DE equation (6) for all $x_0 \in [0, 1]$ also have a design rate (8) of $1-p$ (i.e., it achieves the channel capacity of a BEC whose erasure probability is $p$).

*Proof.* We start with Eq. (6) and proceed by substituting $x_0 = 1 - x$, applying $\lambda^{-1}(\cdot)$ to both sides, and moving things around to get

$$1 - \lambda^{-1}(1 - x) = \left(\frac{1-p}{1-pR(x)}\right)^2 \rho(x). \tag{23}$$

Integrating both sides from $x = 0$ to $x = 1$ gives

$$\int_0^1 \left(1 - \lambda^{-1}(1-x)\right) \, \mathrm{d}x = \int_0^1 \left(\frac{1-p}{1-pR(x)}\right)^2 \rho(x) \, \mathrm{d}x.$$

Since $\lambda(\cdot)$ is positive, monotonic increasing and $\lambda(0) = 0$, $\lambda(1) = 1$, we can use the identity

$$\int_0^1 \lambda(x) \, \mathrm{d}x + \int_0^1 \lambda^{-1}(x) \, \mathrm{d}x = 1 \tag{24}$$

to show that

$$\int_0^1 \lambda(x) \, \mathrm{d}x = \int_0^1 \left(\frac{1-p}{1-pR(x)}\right)^2 \rho(x) \, \mathrm{d}x.$$

Taking the derivative of both sides of Eq. (1) shows that

$$\rho(x) = \int_0^1 \rho(x) \, \mathrm{d}x \cdot R'(x),$$

and then it follows easily that

$$\begin{aligned}
\int_0^1 \lambda(x) \mathrm{d}x &= \int_0^1 \rho(x) \, \mathrm{d}x \cdot \int_0^1 \left(\frac{1-p}{1-pR(x)}\right)^2 R'(x) \mathrm{d}x \\
&= \int_0^1 \rho(x) \, \mathrm{d}x \cdot \int_{R(0)}^{R(1)} \left(\frac{1-p}{1-pu}\right)^2 \mathrm{d}u \\
&= (1-p) \cdot \int_0^1 \rho(x) \, \mathrm{d}x,
\end{aligned}$$

where the fact that $R(0) = 0$ and $R(1) = 1$ is implied by Eq. (1). Dividing both sides by the integral of $\rho(\cdot)$ and using Eq. (8) shows that the design rate $R^{\text{IRA}} = 1 - p$. □

## 4.1 Proof of Theorem 1

### 4.1.1 Finding the D.D. Pair

Consider a bit-regular ensemble of non-systematic IRA codes whose d.d. pair $(\lambda, \rho)$ satisfies the DE equation (6). We approach the problem of finding the d.d. pair by solving Eq. (6) for $\rho(\cdot)$ in terms of $\lambda(\cdot)$ and the progression is actually similar to the proof of Lemma 1, except that the limits of integration change. Starting with Eq. (23) and integrating both sides from $x = 0$ to $x = t$ gives

$$
\begin{aligned}
\int_0^t \left(1 - \lambda^{-1}(1-x)\right) \, dx &= \int_0^t \left(\frac{1-p}{1-pR(x)}\right)^2 \rho(x) \, dx \\
&= \int_0^t \left(\frac{1-p}{1-pR(x)}\right)^2 \frac{R'(x)}{R'(1)} dx \\
&= \frac{(1-p)^2}{R'(1)} \frac{R(t)}{1-pR(t)},
\end{aligned}
\tag{25}
$$

where the substitution $\rho(x) = R'(x)/R'(1)$ follows from Eq. (1). The free parameter $R'(1)$ can be determined by requiring that the d.d. $R(\cdot)$ satisfy $R(1) = 1$. Solving Eq. (25) for $R'(1)$ with $t = 1$ and $R(1) = 1$ shows that

$$
R'(1) = \frac{1-p}{\int_0^1 \left(1 - \lambda^{-1}(1-x)\right) \, dx}.
\tag{26}
$$

Solving Eq. (25) for $R(t)$ and substituting for $R'(1)$ gives

$$
R(t) = \frac{\frac{\int_0^t (1-\lambda^{-1}(1-x)) \, dx}{\int_0^1 (1-\lambda^{-1}(1-x)) \, dx}}{1 - p + p \cdot \frac{\int_0^t (1-\lambda^{-1}(1-x)) \, dx}{\int_0^1 (1-\lambda^{-1}(1-x)) \, dx}}.
$$

For simplicity, we now define

$$
Q(x) \triangleq \frac{\int_0^x \left(1 - \lambda^{-1}(1-t)\right) \, dt}{\int_0^1 \left(1 - \lambda^{-1}(1-t)\right) \, dt},
\tag{27}
$$

substitute $x$ for $t$, and get

$$
R(x) = \frac{Q(x)}{1 - p + pQ(x)}.
\tag{28}
$$

It follows from Eqs. (1) and (28) that

$$
\begin{aligned}
\rho(x) &= \frac{R'(x)}{R'(1)} \\
&= \frac{(1-p)Q'(x)}{(1-p+pQ(x))^2} \cdot \frac{(1-p+pQ(1))^2}{(1-p)Q'(1)} \\
&= \frac{1}{(1-p+pQ(x))^2} \cdot \frac{Q'(x)}{Q'(1)} \\
&= \frac{1 - \lambda^{-1}(1-x)}{(1-p+pQ(x))^2},
\end{aligned}
\tag{29}
$$

and
$$\rho(1) = \frac{1 - \lambda^{-1}(0)}{(1 - p + pQ(1))^2} = 1. \tag{30}$$

The important part of this result is that there is no need to truncate the power series of $\rho(\cdot)$ to force $\rho(1) = 1$. This appears to be an important element of ensembles with bounded complexity.

Now, consider the bit-regular case where every information bit is repeated $q \geq 3$ times (i.e., $\lambda(x) = x^{q-1}$). From Eq. (27), it can be verified with some algebra that

$$Q(x) = qx - (q-1)\left[1 - (1-x)^{\frac{q}{q-1}}\right]. \tag{31}$$

Substituting this into Eq. (29) gives the result in Eq. (10).

Finally, we show that the power series expansion of Eq. (10) defines a proper probability distribution. First, we note that $p \in (0, \frac{1}{13}]$ by hypothesis, and that Appendix A establishes the non-negativity of the d.d. coefficients $\{\rho_n\}$ under this same condition. Since $\rho(1) = 1$, these coefficients must sum to one if the power series expansion converges at $x = 1$. This follows from the asymptotic expansion, given later in (63), which implies that $\rho_n = O(n^{-q/(q-1)})$. Therefore, the function $\rho(x)$ defines a proper d.d.

### 4.1.2 Truncating the D.D.

Starting with the d.d. pair $(\lambda, \rho)$ implied by Eq. (29) (which yields that Eq. (6) holds), we apply Lemma 1 to show that the design rate is $1 - p$. The next step is to slightly modify the check d.d. so that the inequality (7) is satisfied instead. In particular, one can modify the $\rho(x)$ from (29) so that the resulting ensemble of bit-regular non-systematic IRA codes is equal to a fraction $1 - \varepsilon$ of the BEC capacity.

Let us define $M(\varepsilon)$ to be the smallest positive integer $M$ such that the condition in (11) is satisfied. Such an $M$ exists for any $\varepsilon \in (0, 1)$ because $\rho_n = O(n^{-q/(q-1)})$. We define the $\varepsilon$-truncation of $\rho(\cdot)$ to be the new check degree polynomial in (12), which is also equal to

$$\rho_\varepsilon(x) = \left(\rho_1 + \sum_{i=M(\varepsilon)+1}^{\infty} \rho_i\right) + \sum_{i=2}^{M(\varepsilon)} \rho_i x^{i-1}, \tag{32}$$

and satisfies $\rho_\varepsilon(1) = \rho(1) = 1$. Based on Eqs. (11) and (32), and the fact that the power series expansion is non-negative, it is easy to show that

$$\int_0^1 \rho_\varepsilon(x)\,dx < \int_0^1 \rho(x)\,dx + \frac{\varepsilon}{q(1-p)} = \frac{1+\varepsilon}{q(1-p)}.$$

Applying Eq. (8) to the last equation, shows that the design rate of the new ensemble $(\lambda, \rho_\varepsilon)$ of bit-regular, non-systematic IRA codes is given by

$$R^{\text{IRA}} = \frac{\int_0^1 \lambda(x)\,dx}{\int_0^1 \rho_\varepsilon(x)\,dx} = \frac{1}{q \int_0^1 \rho_\varepsilon(x)\,dx} > \frac{1-p}{1+\varepsilon}.$$

Using the fact that $\frac{1}{1+\varepsilon} > 1 - \varepsilon$, for $\varepsilon > 0$, we get the final lower bound

$$R^{\text{IRA}} > (1-p)(1-\varepsilon). \tag{33}$$

This shows that the design rate of the new ensemble of codes is equal at least to a fraction $1 - \varepsilon$ of the capacity of the BEC. Now, we need to show that the new ensemble satisfies the inequality (7), which is required for successful decoding, given by

$$\lambda\left(1 - \left[\frac{1-p}{1 - p\, R_\varepsilon(1-x)}\right]^2 \rho_\varepsilon(1-x)\right) < x, \quad \forall x \in (0, 1] \tag{34}$$

where $R_\varepsilon(\cdot)$ can be computed from $\rho_\varepsilon(\cdot)$ via Eq. (1). Since the truncation of $\rho(x)$ only moves edges from high degree checks (i.e., $x^j$ terms with $j > M$) to degree one checks (i.e. the $x^0$ term), it follows that

$$\rho_\varepsilon(x) > \rho(x), \quad x \in [0, 1). \tag{35}$$

We will also show that

**Lemma 2.**

$$R_\varepsilon(x) > R(x), \quad x \in (0, 1). \tag{36}$$

*Proof.* We rely on Eqs. (1), (10) and (32) to show that for an arbitrary $\varepsilon > 0$

$$R_\varepsilon(x) = \frac{\rho_1 + \sum_{i=M(\varepsilon)+1}^{\infty} \rho_i}{\rho_1 + \sum_{i=M(\varepsilon)+1}^{\infty} \rho_i + \sum_{i=2}^{M(\varepsilon)} \frac{\rho_i}{i}} \cdot x + \frac{\sum_{i=2}^{M(\varepsilon)} \frac{\rho_i}{i} \cdot x^i}{\rho_1 + \sum_{i=M(\varepsilon)+1}^{\infty} \rho_i + \sum_{i=2}^{M(\varepsilon)} \frac{\rho_i}{i}} \triangleq \sum_{i=1}^{\infty} R_i^{(\varepsilon)} x^i \tag{37}$$

and

$$R(x) = \frac{\rho_1}{\rho_1 + \sum_{i=M(\varepsilon)+1}^{\infty} \frac{\rho_i}{i} + \sum_{i=2}^{M(\varepsilon)} \frac{\rho_i}{i}} \cdot x + \frac{\sum_{i=2}^{M(\varepsilon)} \frac{\rho_i}{i} \cdot x^i}{\rho_1 + \sum_{i=M(\varepsilon)+1}^{\infty} \frac{\rho_i}{i} + \sum_{i=2}^{M(\varepsilon)} \frac{\rho_i}{i}} \triangleq \sum_{i=1}^{\infty} R_i\, x^i. \tag{38}$$

It is easy to verify that the coefficients in the power series expansions of $R_\varepsilon(\cdot)$ and $R(\cdot)$ in (37) and (38), respectively, are all non-negative and each of them sum to one. By comparing the two, it follows easily that

$$R_i^{(\varepsilon)} < R_i \quad \forall\, i \geq 2.$$

Since $\sum_{i=1}^{\infty} R_i = \sum_{i=1}^{\infty} R_i^{(\varepsilon)} = 1$, we also see that

$$R_1^{(\varepsilon)} > R_1.$$

Let

$$\delta_i \triangleq R_i - R_i^{(\varepsilon)} \quad i = 1, 2, \ldots$$

then $\delta_1 = -\sum_{i=2}^{\infty} \delta_i$ (since by definition, $\sum_{i=1}^{\infty} \delta_i = \sum_{i=1}^{\infty} R_i - \sum_{i=1}^{\infty} R_i^{(\varepsilon)} = 0$), and $\delta_i > 0$ for every integer $i \geq 2$. It therefore follows that for $x \in (0, 1)$

$$R(x) - R_\varepsilon(x) = \delta_1 x + \sum_{i=2}^{\infty} \delta_i x^i < \delta_1 x + \sum_{i=2}^{\infty} \delta_i x = 0$$

which proves the inequality in (36). □

Finally, the validity of the condition in (34) follows immediately from the two inequalities in (35) and (36), and the fact that the d.d. pair $(\lambda, \rho)$ satisfies the equality in (6) for all $x_0 \in [0, 1]$.

## 4.2 Proof of Theorem 2

### 4.2.1 Finding the D.D. Pair

Like in the proof of Theorem 1, we start the analysis by solving equation (25), but this time we calculate $\lambda(\cdot)$ for a particular choice of $\rho(\cdot)$. Let us choose $\rho(x) = x^2$, so $R(x) = x^3$, and we obtain from (23) that the inverse function of $\lambda(\cdot)$ is equal to

$$\lambda^{-1}(x) = 1 - \left(\frac{1-p}{1-p(1-x)^3}\right)^2 (1-x)^2 . \tag{39}$$

Inserting (39) into (15) shows that the expression of $\lambda(\cdot)$ in (15) is the inverse function to (39) for $x \in [0, 1]$, so (15) gives us a closed form expression of $\lambda(\cdot)$ in the interval $[0, 1]$. As we noted already, for real numbers, one can simplify the expression of $\lambda(\cdot)$ in (15), but since we consider it later as a function of a complex argument, then we prefer to leave it in the form of (15).

In the following, we show how (15) was derived. Note that since we already verified the correctness of (15), then in the following derivation we do not need to worry about issues of convergence. Set

$$y = \lambda^{-1}(x), \quad z = \frac{\sqrt{1-y}}{1-p}, \quad u = 1 - x,$$

Then with these notation and for $u \in [0, 1]$, (39) can be written in the form

$$z\phi(u) = u$$

where $\phi(u) = 1 - pu^3$. We now use the Lagrange inversion formula (see, e.g., [1, Section 2.2]) to obtain the power series expansion of $u = u(z)$ around $z = 0$, i.e., we write

$$u(z) = \sum_{k=0}^{\infty} u_k z^k.$$

If $z = 0$ then $u = 0$, so $u_0 = u(0) = 0$. The Lagrange inversion formula states that

$$u_k = \frac{1}{k} [u^{k-1}] \phi^k(u), \quad k = 1, 2, \ldots \tag{40}$$

where $[u^{k-1}] \phi^k(u)$ is the coefficient of $u^{k-1}$ in the power series expansion of $\phi^k(u)$. From the definition of $\phi(\cdot)$, the binomial formula gives

$$\phi^k(u) = (1 - pu^3)^k = \sum_{j=0}^{k} \left\{ (-1)^j \binom{k}{j} p^j u^{3j} \right\} \tag{41}$$

so from (40) and (41)

$$u_k = \begin{cases} \frac{(-1)^{\frac{k-1}{3}}}{k} \binom{k}{\frac{k-1}{3}} p^{\frac{k-1}{3}}, & \text{if } k = 1, 4, 7, 10, \ldots \\ 0 & \text{otherwise} \end{cases}.$$

We conclude that

$$u(z) = \sum_{k: \frac{k-1}{3} \in \mathbb{N}} \left\{ \frac{(-1)^{\frac{k-1}{3}}}{k} \binom{k}{\frac{k-1}{3}} p^{\frac{k-1}{3}} z^k \right\}$$

where $\mathbb{N}$ designates the set of non-negative integer numbers. From $z = \frac{\sqrt{1-y}}{1-p}$, we get

$$z^k = \frac{(1-y)^{\frac{k}{2}}}{(1-p)^k} = \frac{1}{(1-p)^k} \sum_{n=0}^{\infty} \left\{ \binom{\frac{k}{2}}{n} (-y)^n \right\}$$

so the composition of the two equalities above gives

$$u = \sum_{k:\,\frac{k-1}{3} \in \mathbb{N}} \left\{ \frac{(-1)^{\frac{k-1}{3}}}{k} \binom{k}{\frac{k-1}{3}} \frac{p^{\frac{k-1}{3}}}{(1-p)^k} \sum_{n=0}^{\infty} (-1)^n \binom{\frac{k}{2}}{n} y^n \right\}$$

and $x = 1 - u = \lambda(y)$ (since $y = \lambda^{-1}(x)$). Finally, we obtain a power series expansion for $\lambda(\cdot)$ from the last two equalities

$$\lambda(x) = 1 - \sum_{k:\,\frac{k-1}{3} \in \mathbb{N}} \left\{ \frac{(-1)^{\frac{k-1}{3}}}{k} \binom{k}{\frac{k-1}{3}} \frac{p^{\frac{k-1}{3}}}{(1-p)^k} \cdot \sum_{n=0}^{\infty} (-1)^n \binom{\frac{k}{2}}{n} x^n \right\}.$$

By substituting $k = 3l + 1$ where $l \in \mathbb{N}$, the latter equation can be written as

$$\begin{aligned}
\lambda(x) &= 1 - \frac{1}{\sqrt[3]{p}} \sum_{l=0}^{\infty} \left\{ \frac{(-1)^l}{3l+1} \binom{3l+1}{l} \left(\frac{\sqrt[3]{p}}{1-p}\right)^{3l+1} \sum_{n=0}^{\infty} \binom{\frac{3l+1}{2}}{n} (-x)^n \right\} \\
&= 1 - \frac{1}{\sqrt[3]{p}} \sum_{l=0}^{\infty} \left\{ \frac{(-1)^l}{3l+1} \binom{3l+1}{l} \left(\frac{\sqrt[3]{p}}{1-p}\right)^{3l+1} (1-x)^{\frac{3l+1}{2}} \right\} \\
&= 1 - \frac{1}{\sqrt[3]{p}} \sum_{l=0}^{\infty} \left\{ \frac{(-1)^l}{3l+1} \binom{3l+1}{l} t^{3l+1} \right\}
\end{aligned}$$

where $t \triangleq \left(\frac{\sqrt[3]{p}}{1-p}\right) \sqrt{1-x}$. Fortunately, the final sum can be expressed in closed form and leads to the expression of $\lambda(\cdot)$ in (15). Plots of the function $\lambda(\cdot)$ as a function of $p \in (0,1)$ are depicted in Fig. 2.

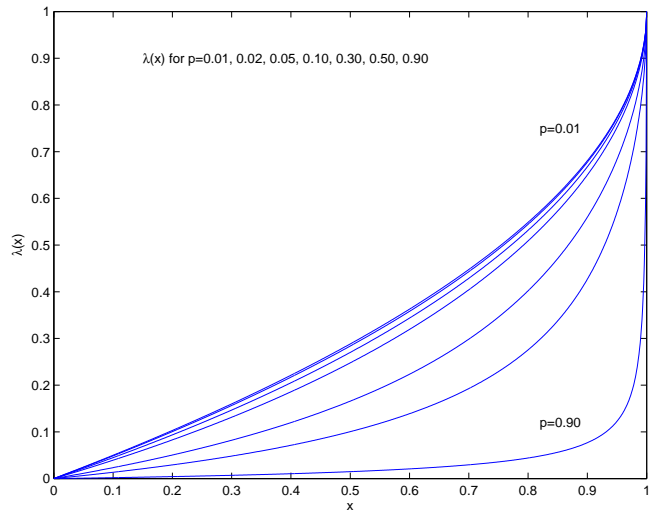

Figure 2: The function $\lambda(\cdot)$ in (15), as a function of the erasure probability $p$ of the BEC.

Finally, we show that the power series expansion of Eq. (15) defines a proper probability distribution. Three different representations of the d.d. coefficients $\{\lambda_n\}$ are presented in Section 5.2.1 and derived in Appendix B. They are also used in Appendix C to prove the non-negativity of the d.d. for $p \in [0, 0.95]$. Since $\lambda(1) = 1$, these coefficients must also sum to one if the power series expansion converges at $x = 1$. The fact that $\lambda_n = O(n^{-3/2})$ follows from a later discussion (in Section 5.2.2) and establishes the power series convergence at $x = 1$. Therefore, the function $\lambda(x)$ gives a well-defined d.d.

### 4.2.2 Truncating the D.D.

Now, we must truncate $\lambda(\cdot)$ in such a way that inequality (7), which is a necessary condition for successful iterative decoding, is satisfied. We do this by treating all information bits with degree greater than some threshold as pilot bits. In practice, this means that the encoder uses a fixed value for each of these bits (usually zero) and the decoder has prior knowledge of these fixed values. This truncation works well because a large number of edges in the decoding graph are initialized by each pilot bit. Since bits chosen to be pilots no longer carry information, the cost of this approach is a reduction in code rate. The rate after truncation is given by

$$R^{\text{IRA}} = \frac{K'}{N} = \frac{K}{N}\frac{K'}{K} = \frac{K}{N}\left(1 - \frac{K-K'}{K}\right),$$

where $N$ is the block length, $K$ is number of information bits before truncation, and $K'$ is the number of information bits after truncation. Applying Lemma 1 to the d.d. pair $(\lambda,\rho)$ shows that the design rate is given by $K/N = 1-p$. Therefore, the rate can be rewritten as $R^{\text{IRA}} = (1-p)(1-\delta)$ where $\delta \triangleq (K - K')/K$ is the fraction of information bits that are used as pilot bits.

For an arbitrary $\varepsilon \in (0, 1)$, we define $M(\varepsilon)$ to be the smallest positive integer $M$ which satisfies Eq. (16). Next, we choose all information bit nodes with degree greater than $M(\varepsilon)$ to be pilot bits. This implies that the fraction of information bit nodes used as pilot bits is given by $\delta = \sum_{n=M(\varepsilon)+1}^{\infty} L_n$ where the fraction of information bit nodes with degree $n$ is given by

$$L_n = \frac{\lambda_n/n}{\sum_{n=2}^{\infty} \lambda_n/n}. \tag{42}$$

Based on Eqs. (8) and (14), we have

$$\sum_{n=2}^{\infty} \frac{\lambda_n}{n} = \int_0^1 \lambda(x)\,dx = R^{\text{IRA}} \int_0^1 \rho(x)\,dx = \frac{1-p}{3}. \tag{43}$$

Therefore, we can use Eqs. (16), (42) and (43) to show that

$$\delta = \sum_{n=M(\varepsilon)+1}^{\infty} L_n = \frac{\sum_{n=M(\varepsilon)+1}^{\infty} \lambda_n/n}{\sum_{n=2}^{\infty} \lambda_n/n} = \frac{\sum_{n=M(\varepsilon)+1}^{\infty} \lambda_n/n}{\frac{1-p}{3}} < \frac{\frac{(1-p)\varepsilon}{3}}{\frac{1-p}{3}} = \varepsilon.$$

Let us define the effective $\varepsilon$-modified d.d. to be

$$\lambda_\varepsilon(x) = \sum_{n=2}^{M(\varepsilon)} \lambda_n x^{n-1}.$$

Although this is not a d.d. in the strict sense (because it no longer sums to one), it is the correct $\lambda$ function for the DE equation. This is because all information bits with degree greater than $M(\varepsilon)$ are known at the receiver and therefore have zero erasure probability. Since the d.d. pair $(\lambda,\rho)$ satisfies the equality in (6) and $\lambda_\varepsilon(x) < \lambda(x)$ for $x \in (0,1]$, then it follows that the inequality required for successful decoding (7) is satisfied.

As explained in Section 1, the encoding and decoding complexity on the BEC are both equal to the number of edges, per information bit, in the Tanner graph. The degree of the parity-check nodes is fixed to 5 (three edges attached to information bits and two edges attached to code bits), and this implies that the complexity is given by

$$\chi_E(\varepsilon,\mathcal{C}) = \chi_D(\varepsilon,\mathcal{C}) = \frac{5}{R^{\text{IRA}}} < \frac{5}{(1-p)(1-\varepsilon)}.$$

Therefore, the complexity is *bounded* and equals $\frac{5}{1-p}$ as the gap to capacity vanishes.

## 4.3 Proof of Theorem 3

*Proof.* Under MPI decoding, the decoding complexity of the sequence of codes $\{\mathcal{C}_m\}$ is equal to the number of edges in the Tanner graph of the original codes $\{\mathcal{C}'_m\}$ normalized per information bit (since for the BEC, one can modify the MPI decoder so that every edge in the Tanner graph is only used once). This normalized number of edges is directly linked to the average degree of the parity-check nodes in the Tanner graphs of the sequence of codes $\{\mathcal{C}'_m\}$ (up to a scaling factor which depends on the rate of the code). We will first derive an information-theoretic bound on the average degree of the parity-check nodes for the sequence $\{\mathcal{C}'_m\}$, say $a_R(\mathcal{C}'_m)$, which will be valid for every decoding algorithm. From this bound, we will directly obtain a bound on the decoding complexity of punctured codes on graphs, when we assume that an MPI decoding algorithm is used.

Let $\mathbf{u}'_m = (u_1, u_2, \ldots, u_{n_m})$ be a codeword of a binary linear block code $\mathcal{C}'_m$, and assume that a subset of the information bits of the code $\mathcal{C}'_m$ are punctured (see footnote no. 5 in p. 7). Let us replace the punctured bits of $\mathbf{u}'_m$ by question marks, and let us call the new vector $\mathbf{u}_m$. The bits of $\mathbf{u}_m$ (those which were not replaced by question marks) are the coordinates of the codewords of the punctured code $\mathcal{C}_m$. Let us assume that $\mathbf{u}_m$ is transmitted over a BEC whose erasure probability is equal to $p$. The question marks in the received vector $\mathbf{v}_m = (v_1, v_2, \ldots, v_{n_m})$ remain in all the places where they existed in $\mathbf{u}_m$ (due to puncturing of a subset of the information bits of $\mathbf{u}'_m$), and in addition, the other bits of $\mathbf{u}_m$ which are transmitted over the BEC are received as question marks with probability $p$ or remain in their original values with probability $1-p$ (due to the erasures of the BEC). Since by our assumption, the sequence of punctured codes $\{\mathcal{C}_m\}$ achieves a fraction $1-\varepsilon$ of the channel capacity with vanishing bit erasure probability, then there exists a decoding algorithm (e.g., ML decoding) so that the average bit erasure probability of the code $\mathcal{C}_m$ goes to zero as we let $m$ tend to infinity, and $\lim_{m\to\infty} R_m \geq (1-\varepsilon)(1-p)$. Here, $R_m$ and $R'_m$ designate the rates (in bits per channel use) of the punctured code $\mathcal{C}_m$ and the original code $\mathcal{C}'_m$, respectively. The rate of the punctured code ($\mathcal{C}_m$) is greater than the rate of the original code ($\mathcal{C}'_m$), i.e., $R'_m < R_m$. Let $P_b^{(i)}(m)$ designate the bit erasure probability of the digit $u_i$ at the end of the decoding process of the punctured code $\mathcal{C}_m$. Without loss of generality, one can assume that the $n_m R'_m$ first bits of the vector $\mathbf{u}'_m$ refer to the information bits of the code $\mathcal{C}'_m$, and the other $n_m(1-R'_m)$ last bits of $\mathbf{u}'_m$ are the parity bits of $\mathcal{C}_m$ and $\mathcal{C}'_m$. Let

$$P_b(m) \triangleq \frac{1}{n_m R'_m} \sum_{i=1}^{n_m R'_m} P_b^{(i)}(m)$$

be the average bit erasure probability of the code $\mathcal{C}_m$ (whose codewords are transmitted with equal probability), based on the observation of the random vector $\mathbf{v}_m$ at the output of the BEC. By knowing the linear block code $\mathcal{C}'_m$, then we get that

$$
\begin{aligned}
\frac{H(\mathbf{u}'_m|\mathbf{v}_m)}{n_m} &= \frac{H(\{u_i\}_{i=1}^{n_m R'_m}|\mathbf{v}_m)}{n_m} + \frac{H(\{u_i\}_{i=n_m R'_m+1}^{n_m}|\mathbf{v}_m, \{u_i\}_{i=1}^{n_m R'_m})}{n_m} \\
&\stackrel{(a)}{=} \frac{H(\{u_i\}_{i=1}^{n_m R'_m}|\mathbf{v}_m)}{n_m} \\
&\stackrel{(b)}{=} \frac{\sum_{i=1}^{n_m R'_m} H(u_i|\mathbf{v}_m, u_1, \ldots, u_{i-1})}{n_m} \\
&\stackrel{(c)}{\leq} \frac{\sum_{i=1}^{n_m R'_m} H(u_i|\mathbf{v}_m)}{n_m} \\
&\stackrel{(d)}{\leq} \frac{\sum_{i=1}^{n_m R'_m} h\left(P_{\text{b}}^{(i)}(m)\right)}{n_m} \\
&\stackrel{(e)}{\leq} R'_m \, h(P_{\text{b}}(m))
\end{aligned}
$$

where equality (a) is valid since the $n_m R'_m$ information bits of the linear block code $\mathcal{C}'_m$ determine the $n_m(1 - R'_m)$ parity bits of its codewords, equality (b) is based on the chain rule for the entropy, inequality (c) follows since conditioning reduces the entropy, inequality (d) follows from Fano's inequality and since the code $\mathcal{C}_m$ is binary, and inequality (e) is based on Jensen's inequality and the concavity of the binary entropy function $h(x) = -x \log_2(x) - (1-x) \log_2(1-x)$ for $x \in (0, 1)$. Based on our assumption that there exists a decoding algorithm so that the *bit erasure probability* of the sequence of codes $\{\mathcal{C}_m\}$ vanishes (as $m \to \infty$), then it follows that

$$\lim_{m \to \infty} \frac{H(\mathbf{u}'_m|\mathbf{v}_m)}{n_m} = 0. \tag{44}$$

For the sake of notational simplicity, we will replace $\mathbf{u}'_m$, $\mathbf{v}_m$ and $n_m$ by $\mathbf{U}'$, $\mathbf{V}$, and $n$, respectively. In the following derivation, let $\mathbf{K}$ and $\mathbf{E}$ designate the random vectors which indicate the positions of the known and punctured/erased digits in the received vector ($\mathbf{V}$), respectively (note that knowing one of these two random vectors implies the knowledge of the other vector). The random vector $\mathbf{V_K}$ denotes the sub-vector of $\mathbf{V}$ with the known digits of the received vector (i.e., those digits which are not punctured by the encoder and not erased by the BEC). Note that there is a *one-to-one* correspondence between the received vector $\mathbf{V}$ and the pair of vectors $(\mathbf{V_K}, \mathbf{E})$. We designate by $\mathbf{U}'_\mathbf{E}$ and $\mathbf{U}'_\mathbf{K}$ the sub-vectors of the original codeword $\mathbf{U}'$ of the code $\mathcal{C}'_m$, such that they correspond to digits of $\mathbf{U}'$ in the punctured/erased and known positions of the received vector, respectively (so that $\mathbf{U}'_\mathbf{K} = \mathbf{V_K}$). Finally, let $H'_\mathbf{E}$ denote the matrix of those columns of $H'$ (a parity-check matrix representing the block code $\mathcal{C}'_m$) whose variables are indexed by $\mathbf{E}$, and $|\mathbf{e}|$ denotes the number of elements of a vector $\mathbf{e}$. Then, we get

$$
\begin{aligned}
H(\mathbf{U}'|\mathbf{V}) &= H(\mathbf{U}'|\mathbf{V_K}, \mathbf{E}) \\
&= H(\mathbf{U}'_\mathbf{E}, \mathbf{U}'_\mathbf{K}|\mathbf{V_K}, \mathbf{E}) \\
&= H(\mathbf{U}'_\mathbf{E}|\mathbf{V_K}, \mathbf{E}) \\
&= \sum_{\mathbf{v_k}, \mathbf{e}} p(\mathbf{v_k}, \mathbf{e}) \, H(\mathbf{U}'_\mathbf{E}|\mathbf{V_K} = \mathbf{v_k}, \mathbf{E} = \mathbf{e}) \\
&= \sum_{\mathbf{v_k}, \mathbf{e}} p(\mathbf{v_k}, \mathbf{e}) \left(|\mathbf{e}| - \text{rank}(H'_\mathbf{e})\right) \\
&= \sum_{\mathbf{e}} p(\mathbf{e}) \left(|\mathbf{e}| - \text{rank}(H'_\mathbf{e})\right) \\
&= \sum_{\mathbf{e}} p(\mathbf{e}) |\mathbf{e}| - \sum_{\mathbf{e}} p(\mathbf{e}) \, \text{rank}(H'_\mathbf{e}) \, .
\end{aligned}
$$

and by normalizing both sides of the equality w.r.t. the block length $(n)$, then

$$\frac{H(\mathbf{U}'|\mathbf{V})}{n} = \frac{1}{n} \sum_{\mathbf{e}} p(\mathbf{e}) |\mathbf{e}| - \frac{1}{n} \sum_{\mathbf{e}} p(\mathbf{e}) \, \text{rank}(H'_\mathbf{e}) \, . \tag{45}$$

Note that the rank of a parity-check matrix $H'_\mathbf{e}$ of the block code $\mathcal{C}'_m$ is upper bounded by the number of non-zero rows of $H'_\mathbf{e}$ which is equal to the number of parity-check nodes which involve punctured or erased bits (the sum $\sum_\mathbf{e} p(\mathbf{e}) \cdot \text{rank}(H'_\mathbf{e})$ is therefore upper bounded by the average number of parity-check sets which involve punctured or erased bits).

Now, we will bound the two sums in the RHS of (45): let $I_m$ and $P_m$ be the number of information bits and parity bits in the original code $\mathcal{C}'_m$. Then $n_m = I_m + P_m$ is the block length of the code $\mathcal{C}'_m$, and the block length of the code $\mathcal{C}_m$ (i.e., the block length after puncturing a fraction $P_{\text{pct}}$ of the information bits in $\mathcal{C}'_m$) is equal to $I_m(1 - P_{\text{pct}}) + P_m$. The rate of the punctured code is therefore equal to $R_m = \frac{I_m}{I_m(1-P_{\text{pct}})+P_m}$. Its asymptotic value (as $m \to \infty$) is by assumption at least $(1-\varepsilon)(1-p)$, i.e.,

$$\lim_{m\to\infty} R_m = \lim_{m\to\infty} \frac{I_m}{I_m(1-P_{\text{pct}})+P_m} \geq (1-\varepsilon)(1-p).$$

We obtain from the last inequality that

$$\lim_{m\to\infty} \frac{I_m}{P_m} \geq \frac{(1-\varepsilon)(1-p)}{P_{\text{eff}} + \varepsilon(1-P_{\text{eff}})}$$

where $P_{\text{eff}}$ was introduced in (20), so the asymptotic rate of the sequence of codes $\{\mathcal{C}'_m\}$ satisfies

$$\lim_{m\to\infty} R'_m = \lim_{m\to\infty} \frac{I_m}{I_m+P_m} \geq \frac{(1-\varepsilon)(1-p)}{(1-\varepsilon)(1-p)+P_{\text{eff}}+\varepsilon(1-P_{\text{eff}})}. \tag{46}$$

The number of elements of a vector $\mathbf{e}$ indicates the number of bits in the codewords of $\mathcal{C}'_m$ which are punctured by the encoder or erased by the BEC. Its average value is therefore equal to

$$\begin{aligned}
\sum_\mathbf{e} p(\mathbf{e}) |\mathbf{e}| &= I_m P_{\text{pct}} + \Big(I_m(1-P_{\text{pct}}) + P_m\Big)p \\
&= I_m P_{\text{pct}} + \frac{I_m p}{R_m} \\
&= n_m R'_m \Big(P_{\text{pct}} + \frac{p}{R_m}\Big),
\end{aligned}$$

so since $R_m < 1 - p$, then we obtain from (46) and the last equality that

$$\lim_{m\to\infty} \frac{1}{n_m} \sum_\mathbf{e} p(\mathbf{e}) |\mathbf{e}| \geq \frac{(1-\varepsilon)(1-p)}{(1-\varepsilon)(1-p)+P_{\text{eff}}+\varepsilon(1-P_{\text{eff}})} \left(P_{\text{pct}} + \frac{p}{1-p}\right). \tag{47}$$

If a parity-check node of the Tanner graph of the code $\mathcal{C}'_m$ is connected to information nodes by $k$ edges, then based on the assumption that the information bits of the codes in $\{\mathcal{C}'_m\}$ are randomly punctured at rate $P_{\text{pct}}$, then the probability that a parity-check node involves at least one punctured or erased information bit is equal to $1 - (1 - P_{\text{eff}})^k$. This expression is valid with probability 1 (w.r.t. the randomly chosen puncturing pattern) when the block length tends to infinity (or in the limit where $m \to \infty$). We note that $P_{\text{eff}}$ is introduced in (20), and it stands for the effective erasure probability of information bits in the code $\mathcal{C}'_m$ when we take into account the effects of the random puncturing of the information bits at the encoder, and the random erasures which are introduced by the BEC. The average number of the parity-check nodes which therefore involve at least one punctured or erased information bit is equal to $n_m(1 - R'_m) \sum_k d_{k,m} \left(1 - (1 - P_{\text{eff}})^k\right)$ where $d_{k,m}$ designates the fraction of parity-check nodes in the Tanner graph of $\mathcal{C}'_m$ which are connected to information nodes by $k$ edges. Therefore

$$\frac{1}{n_m} \sum_\mathbf{e} p(\mathbf{e}) \, \text{rank}(H'_\mathbf{e}) \leq (1 - R'_m)\left(1 - \sum_k d_{k,m}(1 - P_{\text{eff}})^k\right).$$

From Jensen's inequality, we obtain that

$$\sum_k d_{k,m}(1 - P_{\text{eff}})^k \geq (1 - P_{\text{eff}})^{b_R(\mathcal{C}'_m)}$$

where $b_R(\mathcal{C}'_m) \triangleq \sum_k k d_{k,m}$ is the average number of edges which connect a parity-check node with information nodes in the Tanner graph of the code $\mathcal{C}'_m$. By definition, it follows immediately that $a_R(\mathcal{C}'_m) \geq b_R(\mathcal{C}'_m) + l_{\min}$, and therefore we get

$$\frac{1}{n_m} \sum_{\mathbf{e}} p(\mathbf{e}) \operatorname{rank}(H'_{\mathbf{e}}) \leq (1 - R'_m)\left(1 - (1 - P_{\text{eff}})^{a_R(\mathcal{C}'_m) - l_{\min}}\right). \tag{48}$$

From Eqs. (45)–(48), we obtain that

$$\lim_{m \to \infty} \frac{H(\mathbf{u}'_m | \mathbf{v}_m)}{n_m} = \lim_{m \to \infty} \frac{1}{n_m} \sum_{\mathbf{e}} p(\mathbf{e}) |\mathbf{e}| - \lim_{m \to \infty} \frac{1}{n_m} \sum_{\mathbf{e}} p(\mathbf{e}) \operatorname{rank}(H'_{\mathbf{e}})$$

$$\geq \frac{(1-\varepsilon)(1-p)}{(1-\varepsilon)(1-p) + P_{\text{eff}} + \varepsilon(1 - P_{\text{eff}})}\left(P_{\text{pct}} + \frac{p}{1-p}\right)$$

$$- \left(1 - \frac{(1-\varepsilon)(1-p)}{(1-\varepsilon)(1-p) + P_{\text{eff}} + \varepsilon(1 - P_{\text{eff}})}\right)\left(1 - (1 - P_{\text{eff}})^{a_R - l_{\min}}\right)$$

where $a_R \triangleq \liminf_{m \to \infty} a_R(\mathcal{C}'_m)$. The limit of the normalized conditional entropy in (44) is equal to zero, so its lower bound in the last inequality cannot be positive. This yields the inequality

$$(1-p)(1-\varepsilon)P_{\text{pct}} + (1-\varepsilon)p - \left(P_{\text{eff}} + \varepsilon(1 - P_{\text{eff}})\right)\left(1 - (1 - P_{\text{eff}})^{a_R - l_{\min}}\right) \leq 0. \tag{49}$$

From (20), then $(1-p) P_{\text{pct}} + p - P_{\text{eff}} = 0$, so simplification of the LHS in (49) gives

$$P_{\text{eff}} (1 - P_{\text{eff}})^{a_R - l_{\min}} \leq \varepsilon(1-p)P_{\text{pct}} + \varepsilon p + \varepsilon(1 - P_{\text{eff}})\left(1 - (1 - P_{\text{eff}})^{a_R - l_{\min}}\right)$$

$$\leq \varepsilon(1-p)P_{\text{pct}} + \varepsilon p + \varepsilon(1 - P_{\text{eff}})$$

$$= \varepsilon.$$

This yields the following information-theoretic bound the asymptotic degree of the parity-check nodes ($a_R(\mathcal{C}'_m)$)

$$\liminf_{m \to \infty} a_R(\mathcal{C}'_m) \geq \frac{\ln\left(\frac{P_{\text{eff}}}{\varepsilon}\right)}{\ln\left(\frac{1}{1-P_{\text{eff}}}\right)} + l_{\min}, \tag{50}$$

which is valid with probability 1 w.r.t. the puncturing patterns. The proof until now is valid under any decoding algorithm (even the optimal MAP decoding algorithm), and in the continuation, we link our result to MPI decoding.

From the information-theoretic bound in (50), it follows that with probability 1 w.r.t. the puncturing patterns, the asymptotic decoding complexity of the sequence of punctured codes $\{\mathcal{C}_m\}$ satisfies under MPI decoding

$$\liminf_{m \to \infty} \chi_D(\mathcal{C}_m) = \left(\frac{1-R}{R}\right) \liminf_{m \to \infty} a_R(\mathcal{C}'_m)$$

where $R$ is the asymptotic rate of the sequence $\{\mathcal{C}_m\}$. The scaling by $\frac{1-R}{R}$ is due to the fact that the complexity is (by definition) normalized per information bit, and the average degree of the check nodes is normalized per parity-check node). As said before, the last equality is true since the MPI decoder can be modified for a BEC so that every edge in the Tanner graph is only used once; therefore, the number of operations which are performed for MPI decoding of the punctured code $\mathcal{C}_m$ is equal to the number of edges in the Tanner graph of the original code $\mathcal{C}'_m$. Since $R \leq 1 - p$, then we obtain from (50) that under MPI decoding, the asymptotic decoding complexity satisfies (19) with probability 1 w.r.t. the puncturing patterns.

If $P_{\text{pct}} = 1 - O(\varepsilon)$, then it follows from (20) that also $P_{\text{eff}} = 1 - O(\varepsilon)$. Therefore, the RHS of (19) remains bounded when the gap (in rate) to capacity vanishes (i.e., in the limit where $\varepsilon \to 0$). We conclude that with probability 1 w.r.t. the puncturing patterns of the information bits, a necessary condition that the sequence of punctured codes achieves the capacity of the BEC with *bounded complexity* under MPI decoding is that the puncturing rate of the information bits satisfies the condition $P_{\text{pct}} = 1 - O(\varepsilon)$. Otherwise, the complexity grows like $O\left(\ln\left(\frac{1}{\varepsilon}\right)\right)$. □

*Discussion:* Note that a-fortiori the same statement in Theorem 3 holds if we require that the block erasure probability tends asymptotically to zero. We note that this statement is valid for *every sequence* of codes, as opposed to a lower bound on the decoding complexity of IRA ensembles on the BEC which we originally derived based on the density evolution (DE) equation. Considering ensembles of non-systematic IRA codes with random puncturing of the information bits, the lower bound on the complexity that we derived from the DE equation (based on a natural generalization of the derivation of the bound in [15, Theorem 1] to the case of randomly punctured IRA code ensembles) was a slightly looser bound than the bound in Theorem 3, so we omit its derivation. The lower bound on decoding complexity of capacity-achieving codes on the BEC is especially interesting due to two constructions of capacity-achieving IRA ensembles on the BEC with bounded complexity that were introduced in the first two theorems of our paper. We note that for ensembles of IRA codes where the inner code is a differential encoder, together with the choice of puncturing systematic bits of the IRA codes and the natural selection of the information bits as the systematic bits, then $l_{\min} = 2$ (since every parity-check node is connected to exactly two parity bits). For punctured IRA codes, the lower bound in (19) is also a lower bound on the encoding complexity (since the encoding complexity of IRA codes is equal to the number of edges in the Tanner graph per information bit, so under MPI decoding, the encoding and the decoding complexity of IRA codes on the BEC are the same).

The lower bound on the asymptotic degree of the parity-check nodes in (50) is valid under ML decoding (and hence, it is also valid under any sub-optimal decoding algorithm, such as MPI decoding). Finally, the link between the degree of the parity-check nodes in the Tanner graph and the decoding complexity is valid under MPI decoding.

## 4.4 Proof of Theorem 4

*Proof.* The proof relies on the proofs of [2, Theorem 1] and [14, Theorem 1], and it suggests a generalization to the case where a fraction of the information bits are punctured before the code is transmitted over an MBIOS channel.

Under MPI decoding, the decoding complexity per iteration of the sequence of codes $\{\mathcal{C}_m\}$ is equal to the number of edges in the Tanner graph of the original codes $\{\mathcal{C}'_m\}$ normalized per information bit. Similarly to the proof for the BEC, we will first derive an information-theoretic bound on the average degree of the parity-check nodes for the sequence $\{\mathcal{C}'_m\}$, say $a_R(\mathcal{C}'_m)$, which will be valid for every decoding algorithm. From this bound, we will directly obtain a bound on the decoding complexity per iteration of punctured codes on graphs, when we assume that an MPI decoding algorithm is performed.

It suffices to prove the first bound (which refers to the limit of the average degree of the parity-check nodes for the sequence $\{\mathcal{C}'_m\}$) w.r.t. MAP decoding. This is because the MAP algorithm minimizes the bit error probability and therefore achieves at least the same fraction of capacity as any suboptimal decoding algorithm. According to our assumption about random puncturing of the information bits at rate $P_{\text{pct}}$, then it follows that the equivalent MBIOS channel for the information bits is given by

$$q(y|x=1) = P_{\text{pct}}\, \delta_0(y) + (1 - P_{\text{pct}})\, p(y|x=1) \tag{51}$$

which is physically degraded w.r.t. the original communication channel whose conditional *pdf* (given that $x=1$ is the input to the channel) is $p(y|x=1)$. On the other hand, since we assume that only information bits are punctured, then the original MBIOS channel over which the communication takes place is also the equivalent channel for the parity bits.

By assumption, the sequence of punctured codes $\{\mathcal{C}_m\}$ achieves a fraction $1 - \varepsilon$ of the channel capacity. Let $I_m$ and $P_m$ designate the number of information bits and parity bits in the code $\mathcal{C}'_m$ (before puncturing), then the rate of the punctured code $\mathcal{C}_m$ is given by

$$R_m = \frac{I_m}{(1 - P_{\text{pct}})I_m + P_m}\,.$$

According to the assumption in the theorem, we have $\lim_{m \to \infty} R_m \geq (1-\varepsilon)C$, which implies that

$$\lim_{m \to \infty} \frac{I_m}{P_m} \geq \frac{(1-\varepsilon)C}{1 - (1-\varepsilon)(1-P_{\text{pct}})C}\,. \tag{52}$$

The asymptotic rate of the original sequence of codes $\{\mathcal{C}'_m\}$ (before puncturing) therefore satisfies

$$\lim_{m \to \infty} R'_m = \lim_{m \to \infty} \frac{I_m}{I_m + P_m} \geq \frac{(1-\varepsilon)C}{1 + (1-\varepsilon)P_{\text{pct}}C} \tag{53}$$

Similarly to the information-theoretic proof for the BEC, it follows exactly via the same chain of inequalities that

$$\lim_{m \to \infty} \frac{H(\mathbf{u}'_m | \mathbf{v}_m)}{n_m} = 0 \tag{54}$$

where $\mathbf{u}'_m$ and $\mathbf{v_m}$ designate a codeword of the original code $\mathcal{C}'_m$, and the received vector at the output of the channel (after puncturing information bits from $\mathbf{u}'_m$ and transmitting the punctured codeword over the communication channel), respectively. The parameter $n_m$ designates the block length of the original code $\mathcal{C}'_m$.

Let $g(y|x=1)$ be an arbitrary conditional *pdf* at the output of an MBIOS channel, given that $x=1$ is the input to this channel, and let us define the operator

$$\omega(g) \triangleq \frac{1}{2} \int_{-\infty}^{+\infty} \min\bigl(g(y|x=1), g(y|x=0)\bigr)\, dy.$$

Then, it follows directly that $0 \leq \omega(g) \leq \frac{1}{2}$, and $\omega(p) = w$ where $w$ is introduced in (22). For the MBIOS channel in (51)

$$\begin{aligned}
\omega(q) &= \frac{P_{\text{pct}}}{2} \int_{-\infty}^{+\infty} \delta_0(y)\, dy + \frac{1 - P_{\text{pct}}}{2} \int_{-\infty}^{+\infty} \min\bigl(p(y|x=1), p(y|x=0)\bigr)\, dy \\
&= \frac{P_{\text{pct}}}{2} + (1 - P_{\text{pct}})\, w
\end{aligned}$$

so, we obtain the equality

$$1 - 2\omega(q) = (1 - 2w)(1 - P_{\text{pct}}). \tag{55}$$

Analogously to the proof of [2, Theorem 1], let us define a binary random vector $\mathbf{Z} = (z_1, \ldots, z_{n_m})$ so that for $l = 1, 2, \ldots, n_m$, if $u_l$ and $v_l$ designate the $l$-th components of $\mathbf{u}'_m$ and $\mathbf{v}_m$, respectively, then

$$\Pr\bigl(z_l = 1 \bigm| q(v_l|u_l = 1) > q(-v_l|u_l = 1)\bigr) = 1$$

$$\Pr\bigl(z_l = 0 \bigm| q(v_l|u_l = 1) < q(-v_l|u_l = 1)\bigr) = 1$$

$$\Pr\bigl(z_l = 1 \bigm| q(v_l|u_l = 1) = q(-v_l|u_l = 1)\bigr) = \frac{1}{2}.$$

In particular, in case that $v_l$ corresponds to an erasure, then $z_l$ is equal to zero or one with probability $\frac{1}{2}$. Hence, the channel $\mathbf{U}' \to \mathbf{Z}$ is equivalent to a BSC with crossover probability which is equal to $\omega(q)$.

Based on [2, Eqs. (4), (5), (11) and (12)], we obtain that

$$\frac{H(\mathbf{u}'_m|\mathbf{v}_m)}{n_m} \geq 1 - \frac{I(\mathbf{u}'_m; \mathbf{v}_m)}{n_m} - \frac{H(\mathbf{S}'_m)}{n_m} \tag{56}$$

where $\mathbf{S}'_m = H'_m \mathbf{Z}^T$ is the syndrome (we designate by $H'_m$ a parity-check matrix of the code $\mathcal{C}'_m$). From [2, Eq. (14)], we obtain an upper bound on the normalized entropy of the syndrome for the case of random puncturing

$$\frac{H(\mathbf{S}'_m)}{n_m} \leq (1 - R'_m)\, h\!\left(\frac{1 - (1 - 2\omega(q))^{a_R(\mathcal{C}'_m)}}{2}\right) \tag{57}$$

where as compared to [2, Eq. (14)], we further loosen the upper bound on the entropy of the syndrome by assuming that all the bits of $\mathcal{C}'_m$ face (because of puncturing) the channel in (51). In fact, this is true only for the information bits of the codewords of $\mathcal{C}'_m$ (since parity bits are not punctured), but since the channel with the conditional *pdf* $q(\cdot|x = 1)$ is physically degraded w.r.t. to the original MBIOS channel, and the inequality $\omega(q) \geq w$ follows directly from (55), then the upper bound in (57) holds.

Let $L_m$ designate the number of digits of the codewords in $\{\mathcal{C}_m\}$ (i.e., after puncturing information bits at a puncturing rate $P_{\text{pct}}$), then $L_m = I_m(1 - P_{\text{pct}}) + P_m$, and

$$\frac{I(\mathbf{u}'_m; \mathbf{v}_m)}{L_m} \leq C \tag{58}$$

and from (52), we get

$$\lim_{m \to \infty} \frac{L_m}{n_m} = \lim_{m \to \infty} \frac{\frac{I_m}{P_m} \cdot (1 - P_{\text{pct}}) + 1}{\frac{I_m}{P_m} + 1} \leq \frac{1}{1 + (1 - \varepsilon) P_{\text{pct}} C}. \tag{59}$$

Since from (54), the LHS of (56) vanishes as we let $m$ tend to infinity, then the combination of Eqs. (52)–(59) gives the following inequality in the limit where $m \to \infty$

$$1 - \frac{C}{1 + (1 - \varepsilon) P_{\text{pct}} C} - \left(1 - \frac{(1 - \varepsilon) C}{1 + (1 - \varepsilon) P_{\text{pct}} C}\right) h\!\left(\frac{1 - (1 - 2\omega(q))^{a_R}}{2}\right) \leq 0$$

where $a_R \triangleq \lim_{m \to \infty} a_R(\mathcal{C}'_m)$. By invoking the following inequality for the binary entropy function (see [14, Lemma 3.1 (p. 1618)])

$$h(x) \leq 1 - \frac{2}{\ln 2}\left(x - \frac{1}{2}\right)^2, \quad 0 \leq x \leq \frac{1}{2},$$

we obtain from the last two inequalities that

$$1 - \frac{C}{1 + (1-\varepsilon)P_{\text{pct}}C} - \left(1 - \frac{(1-\varepsilon)C}{1 + (1-\varepsilon)P_{\text{pct}}C}\right)\left(1 - \frac{(1 - 2\omega(q))^{2a_R}}{2\ln 2}\right) \leq 0.$$

We note that the last inequality extends the inequality in [14, Eq. (32)] to the case where we allow random puncturing of the information bits at an arbitrary puncturing rate (if there is no puncturing, then $P_{\text{pct}} = 0$ and $\omega(q) = w$ from (55), so [14, Eq. (32)] follows directly as a particular case). From the last inequality and (55), we obtain that with probability 1 w.r.t. the puncturing patterns, the following information-theoretic bound on the asymptotic degree of the parity-check nodes is satisfied

$$\liminf_{m \to \infty} a_R(\mathcal{C}'_m) \geq \frac{\ln\left(\frac{1}{\varepsilon} \frac{1 - (1-P_{\text{pct}})C}{2C \ln 2}\right)}{2\ln\left(\frac{1}{(1-P_{\text{pct}})(1-2w)}\right)} . \tag{60}$$

The proof until now is valid under an arbitrary decoding algorithm (i.e., under MAP decoding, and hence, under any other decoding algorithm). In order to proceed, we refer to MPI decoding, where the asymptotic decoding complexity per iteration of the punctured codes $\mathcal{C}_m$ is equal to $\frac{1-R}{R}$ times $a_R$ where $R \triangleq \lim_{m \to \infty} R_m$ is the asymptotic rate of the sequence of punctured codes $\{\mathcal{C}_m\}$. By assumption $R \geq (1-\varepsilon)C$, so we obtain from (60) the lower bound on the decoding complexity per iteration of punctured codes which is given in (21). This lower bound drives us to the interesting conclusion that if the puncturing rate of the information bits is strictly less than 1, then the decoding complexity per iteration must grow at least like $\ln\left(\frac{1}{\varepsilon}\right)$. On the other hand, if $P_{\text{pct}} = 1 - O(\varepsilon)$, then the numerator and denominator of the RHS in (21) are both in the order of $\ln\left(\frac{1}{\varepsilon}\right)$, and therefore the lower bound on the decoding complexity per iteration stays bounded as the gap (in rate) to capacity vanishes. $\square$

*Discussion:* Like Theorem 3, the lower bound on the decoding complexity in Theorem 4 also clearly holds if we require vanishing block error probability. For a BEC with erasure probability $p$, the parameter $w$ in (22) is equal to $w = \frac{p}{2}$, and the capacity is equal to $C = 1 - p$. From (20), this implies that for the BEC, $(1-2w)(1-P_{\text{pct}}) = 1 - P_{\text{eff}}$. It therefore follows that the lower bound in Theorem 3 is at least twice larger than the lower bound for the BEC which we get from the general bound in Theorem 4. The derivation of Theorems 3 and 4 are also different, so because of these two reasons, we derive the stronger version of the bound for the BEC in addition to the general bound for MBIOS channels. The comparison between the two bounds for punctured codes on graphs is consistent with the parallel comparison in [14, Theorem 1] for non-punctured codes. The bounds in Theorems 3 and 4 refer both to *random puncturing*, so these two theorems are valid with probability 1 w.r.t. the puncturing patterns, if we let the block length of the codes go to infinity. Since these bounds become trivial for some deterministic puncturing patterns, it remains an interesting open problem to derive information-theoretic bounds that can be applied to *every puncturing pattern*.

# 5 Analytical Properties and Efficient Computation of the D.D.

In this section, we tackle the problem of computing the d.d. coefficients for both the bit-regular and check-regular ensembles. While doing this, we also lay some of the groundwork required to prove that these coefficients are non-negative. Asymptotic expressions for the coefficients can also be computed rather easily in the process. We start with bit-regular ensemble because the analysis is somewhat simpler.

## 5.1 The Bit-Regular Ensemble

### 5.1.1 A Recursion for the D.D. Coefficients $\{\rho_n\}$

We present here an efficient way for the calculation of the d.d. coefficients $\{\rho_n\}$, referring to the ensemble of bit-regular IRA codes in Theorem 1. To this end, we derive a very simple recursion for coefficients of the power series expansions of $R(x)$ and $\rho(x)$. We start with Eq. (28) and rearrange things to get

$$R(x) = \frac{1}{1-p} Q(x) - \frac{p}{1-p} R(x)Q(x).$$

Next, we substitute the power series expansions for $R(x)$ and $Q(x)$ to get

$$\sum_{n=2}^{\infty} R_n x^n = \frac{1}{1-p} \sum_{n=2}^{\infty} Q_n x^n - \frac{p}{1-p} \sum_{i=2}^{\infty} R_i x^i \sum_{j=2}^{\infty} Q_j x^j.$$

Matching the coefficients of $x^n$ on both sides gives

$$R_n = \frac{1}{1-p} Q_n - \frac{p}{1-p} \sum_{i=2}^{n-2} R_i Q_{n-i}. \tag{61}$$

Using Eq. (27), we can write

$$\begin{aligned} Q(x) &= \frac{\int_0^x (1-(1-t)^{1/(q-1)}) \mathrm{d}t}{\int_0^1 (1-(1-t)^{1/(q-1)}) \mathrm{d}t} \\ &= \frac{\int_0^x \sum_{k=1}^{\infty} (-1)^{k+1} \binom{\frac{1}{q-1}}{k} t^k \mathrm{d}t}{\frac{1}{q}} \\ &= q \sum_{k=1}^{\infty} \binom{\frac{1}{q-1}}{k} \frac{(-x)^{k+1}}{k+1}, \end{aligned}$$

which implies that

$$Q_n = \frac{(-1)^n q}{n} \binom{\frac{1}{q-1}}{n-1}, \quad n \geq 2. \tag{62}$$

From (61) and (62), it follows that

$$R_n = \frac{(-1)^n q}{n(1-p)} \binom{\frac{1}{q-1}}{n-1} - \frac{pq}{1-p} \sum_{i=2}^{n-2} R_i \frac{(-1)^{n-i}}{n-i} \binom{\frac{1}{q-1}}{n-i-1}, \quad n \geq 2.$$

Since $\rho(x) = R'(x)/R'(1)$, this gives

$$\rho_n = \frac{n\,R_n}{R'(1)} = \frac{n\,R_n}{q\,(1-p)}$$

where the last equality follows from Eqs. (9) and (26).

### 5.1.2 Asymptotic Behavior of $\rho_n$

In this section, we consider the asymptotic behavior of the coefficients in the power series expansion of Eq. (10). The resulting expression provides important information about the decay rate of the coefficients. It also shows that $\rho_n$ always becomes positive for large enough $n$ and therefore lends support to Conjecture 1. We approach the problem by first writing $\rho(x)$ as a power series in $(1-x)^{1/(q-1)}$ and then analyzing the asymptotic behavior of each term. This approach is motivated and justified by the results of [3].

We start by rewriting Eq. (10) in terms of $u = (1-x)^{1/(q-1)}$, and then expanding the result into a power series in $u$ to get

$$\begin{aligned}\rho(x) &= \frac{1-u}{\left(1 - p\big(qu^{q-1} - (q-1)u^q\big)\right)^2} \\ &= (1-u)\sum_{i=0}^{\infty}(i+1)p^i\left(qu^{q-1} - (q-1)u^q\right)^i \\ &= 1 - u + 2pqu^{q-1} - 2p(2q-1)u^q + 2p(q-1)u^{q+1} + 3p^2q^2u^{2q-2} + O\left(u^{2q-1}\right).\end{aligned}$$

Now, we can convert this into an asymptotic estimate of $\rho_n$ by using the fact (from [3]) that

$$[x^n](1-x)^\alpha = \frac{n^{-1-\alpha}}{\Gamma(-\alpha)}\left(1 + \frac{\alpha(\alpha+1)}{2n} + O\left(n^{-2}\right)\right),$$

where $[x^k]A(x)$ is the coefficient of $x^k$ in the power series expansion of $A(x)$. We note that the terms $u^{q-1}$ and $u^{2q-2}$ are actually polynomials in $x$ and do not contribute to the asymptotic behavior. Combining the remaining terms with the equality $\rho_{n+1} = [x^n]\rho(x)$ shows that

$$\rho_{n+1} = \frac{n^{-\frac{q}{q-1}}}{(q-1)\Gamma(\frac{q-2}{q-1})}\left(1 + \frac{q}{2(q-1)^2 n} - \frac{2pq(2q-1)}{(q-1)n} + \frac{4p(q+1)\Gamma(\frac{q-2}{q-1})}{\Gamma(\frac{q-3}{q-1})n^{\frac{q}{q-1}}} + O\left(n^{-2}\right)\right). \tag{63}$$

We note that Gamma functions with negative arguments were modified to have positive arguments using the identity $\Gamma(x) = \Gamma(x+1)/x$.

## 5.2 The Check-Regular Ensemble

### 5.2.1 Three Representations for the D.D. Coefficients $\{\lambda_n\}$

We present here three different useful expressions for the computation of the d.d. coefficients $\{\lambda_n\}$, referring to the ensemble of check-regular IRA codes in Theorem 2. The first expression is based on the Lagrange inversion formula (see, e.g., [1, Section 2.2]). The second expression follows by applying the Cauchy residue theorem in order to obtain the power series expansion of $\lambda(\cdot)$ in (15), and the third expression provides a simple recursion which follows from the previous expression. These three expressions are proved in Appendix B, and provide efficient numerical methods to calculate the d.d. of the information bits (from the edge perspective).

**First expression:**

$$\lambda_n = \frac{1}{n-1}[x^{n-2}]\phi^{n-1}(x), \qquad n = 2, 3, \ldots \tag{64}$$

where
$$\phi(x) \triangleq \frac{x}{1 - \left(\frac{1-p}{1-p(1-x)^3}\right)^2 (1-x)^2} . \tag{65}$$

This representation of the sequence $\{\lambda_n\}$ follows directly from the Lagrange inversion formula by writing $y\phi(x) = x$ where $y = \lambda^{-1}(x)$.

**Second expression:**

$$\lambda_n = -\frac{4(1-p)^2}{9\pi p} \operatorname{Im} \left\{ \int_0^\infty \frac{g(r)}{\left(1 + c(p)\, re^{\frac{i\pi}{3}}\right)^n} \, dr \right\}, \quad n = 2, 3, \ldots \tag{66}$$

where
$$c(p) \triangleq \left(\frac{4}{27} \frac{(1-p)^3}{p}\right)^{\frac{2}{3}}, \quad 0 < p < 1 \tag{67}$$

and the complex function $g(\cdot)$ is

$$g(r) = \lim_{w \to 0^+} \frac{\sin\left(\frac{1}{3} \arcsin(r^{\frac{3}{4}} + iw)\right)}{r^{\frac{1}{4}}}, \quad r > 0. \tag{68}$$

**Third expression:**

$$\lambda_{n+1}(p) = \frac{1-p}{(1+2p)^{2n-1}} \cdot \sum_{i=0}^{2(n-1)} a_i^{(n)} p^i, \quad n = 1, 2, \ldots \tag{69}$$

where for $n \geq 2$, we calculate the coefficients $\left\{a_i^{(n)}\right\}_{i=0}^{2(n-1)}$ with the recursive equation

$$a_i^{(m+1)} = \frac{(2m - 3i - 1)\left(a_i^{(m)} - 2a_{i-2}^{(m)}\right) + (20m - 3i - 1)a_{i-1}^{(m)}}{2(m+1)}, \quad m = 1, 2, \ldots, n-1 \tag{70}$$

and the initial condition $a_0^{(1)} = \frac{1}{2}$. We define $a_i^{(k)}$ as zero for $i < 0$ or $i > 2(k-1)$ (where $k = 1, 2, \ldots$). Based on Eqs. (69) and (70), it follows that

$$\lambda_2(p) = \frac{1-p}{2(1+2p)}$$

$$\lambda_3(p) = \frac{(1-p)(1 + 16p + 10p^2)}{8(1+2p)^3}$$

$$\lambda_4(p) = \frac{(1-p)(1 + 12p + 168p^2 + 164p^3 + 60p^4)}{16(1+2p)^5}$$

$$\lambda_5(p) = \frac{(1-p)(5 + 80p + 470p^2 + 7840p^3 + 9640p^4 + 5920p^5 + 1560p^6)}{128(1+2p)^7}$$

and so on. Eqs. (69) and (70) provide a simple way to calculate the sequence $\{\lambda_n\}$ without the need to calculate numerically complicated improper integrals or to obtain the power series expansion of a complicated function; this algorithm involves only the four elementary operations, and hence can be implemented very easily.

An alternative way to express Eqs. (69) and (70) is

$$\lambda_{n+1}(p) = \frac{(1-p) \cdot P_n(p)}{(1+2p)^{2n-1}} \qquad n = 1, 2, \ldots \qquad (71)$$

where $\{P_n(x)\}_{n \geq 1}$ is a sequence of polynomials of degree $2(n-1)$ which can be calculated with the recursive equation

$$P_{n+1}(x) = \frac{\left[(14 - 4n)x^2 + (20n - 4)x + 2n - 1\right] P_n(x) - 3x(1 + x - 2x^2) \frac{\mathrm{d}P_n(x)}{\mathrm{d}x}}{2(n+1)} \qquad n = 1, 2, \ldots \qquad (72)$$

and the initial polynomial $P_1(x) = \frac{1}{2}$.

### 5.2.2  Asymptotic Behavior of $\lambda_n$

We show in Appendix B that for a fixed value of $p$, the asymptotic behavior of the d.d. $\{\lambda_n\}$ is given by

$$\lambda_{n+1} = \frac{n^{-\frac{3}{2}}}{2\sqrt{\pi}(1-p)} \left[1 + \frac{3\sqrt{2}}{2} a_p^{-(n-\frac{1}{2})} \sin\left(\left(n - \frac{1}{2}\right)\theta_p\right)\right]\left[1 + \frac{3}{8n} + \frac{25}{128n^2} + O\left(\frac{1}{n^3}\right)\right] \qquad (73)$$

where

$$a_p \triangleq \left|1 + e^{\frac{i\pi}{3}} c(p)\right|, \qquad \theta_p \triangleq \arg\left[1 + e^{\frac{i\pi}{3}} c(p)\right] \qquad (74)$$

and $c(p)$ is given in (67). Unless $p$ is close to one, it can be verified that the asymptotic expression for the coefficients $\{\lambda_n\}$ also provides a tight approximation for these coefficients already for moderate values of $n$. For example, if $p = 0.5$, then the asymptotic expression in (73) is tight for $n \geq 20$ ($\lambda_{20}$ is equal to 0.0100 while the asymptotic expression in (73) is equal to 0.0107). If $p = 0.8$, the asymptotic expression for the coefficients $\{\lambda_n\}$ in (73) is tight only for $n \geq 120$ (the approximation for $\lambda_{120}$ is equal to 0.0020, and its exact value is 0.0021). In general, by increasing the value of $p$ (which is less than unity), the asymptotic expression in (73) becomes a good approximation for the coefficients $\lambda_n$ starting from a higher value of $n$ (see Fig. 3).

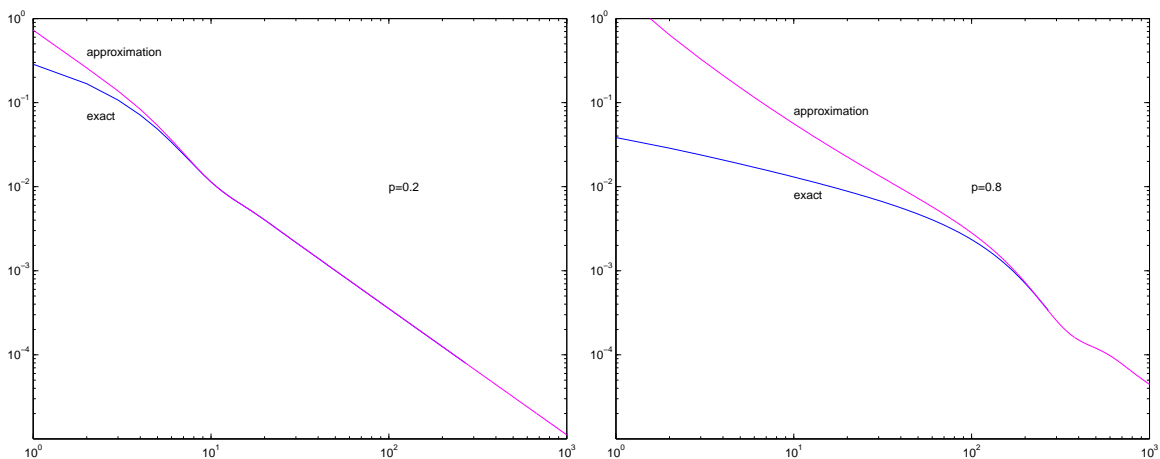

Figure 3: The exact and approximated values of $\lambda_{n+1}$ ($n = 1, \ldots, 1000$) for the check-regular IRA ensemble in Theorem 2. The approximation is based on the asymptotic expression in (73). The plots refer to a BEC whose erasure probability is $p = 0.2$ (left plot) and $p = 0.8$ (right plot).

### 5.2.3 Some Properties of the Power Series Expansion of $\lambda(x)$ in Eq. (15)

The values of $P_n(\cdot)$ at the endpoints of the interval $[0,1]$ can be calculated from the recursive equation (72) (the coefficient of the derivative vanishes at these endpoints). Calculation shows that

$$P_n(0) = \frac{1}{2n-1}\frac{1}{4^n}\binom{2n}{n}, \quad P_n(1) = \frac{9^{n-1}}{4^n}\binom{2n}{n} \quad n=1,2,\ldots \tag{75}$$

Since these values are positive, it follows from (71) that $\{\lambda_n(p)\}_{n\geq 2}$ are positive for $0 \leq p < 1$ if and only if the polynomials $P_n(\cdot)$ do not have zeros inside the interval $[0,1]$ for all $n \geq 1$. We note that if for every $n \geq 1$, all the coefficients of the polynomial $P_n(\cdot)$ were positive, then based on (71), this could suggest a promising direction to prove the positivity of $\{\lambda_n(p)\}_{n\geq 2}$ over the whole interval $p \in [0,1)$. Unfortunately, this is only true for $n \leq 6$. The positivity of $\{\lambda_n(p)\}_{n\geq 2}$ over the interval $[0, 0.95]$ is proved in Appendix C, based on their relation to the polynomials $P_n(\cdot)$. As we already noted, our numerical results strongly support the conjecture that this is true also for $p \in (0.95, 1)$.

As we will see, the behavior of the functions $\lambda_n(p)$, in the limit where $p \to 1$ and $n \to \infty$, depends on the order that the limits are taken. When the value of $n$ is fixed, it follows from (71) and (75) that in the limit where $p \to 1$

$$\lambda_n(p) \approx \frac{P_{n-1}(1)}{3^{2n-3}}(1-p) = \frac{1}{3}\frac{1}{4^{n-1}}\binom{2n-2}{n-1}(1-p) \quad n=2,3,4,\ldots \tag{76}$$

Therefore, for a fixed value of $n$, $\lambda_n(p)$ is linearly proportional to $1-p$ when $p \to 1$. On the other hand, if the value of $p$ is fixed ($0 < p < 1$) and we let $n$ tend to infinity, then we obtain from (73) that $\lambda_n(p)$ is inversely proportional to $1-p$. Observe from Fig. 3 that the sequence of functions $\{\lambda_n(p)\}_{n\geq 2}$ is monotonically decreasing for all values of $p$, and that the tail of this sequence becomes more significant as the value of $p$ grows. This phenomenon can be explained by Eq. (73) since the asymptotic behavior of the sequence $\{\lambda_n(p)\}$ is linearly proportional to $\frac{1}{1-p}$; this makes the tail of this sequence more significant as the value of $p$ is closer to 1. It seems from the plots in Fig. 3 that the asymptotic expression (73) forms an upper bound on the sequence $\{\lambda_n(p)\}$. From these plots, it follows that by increasing the value of $p$, then the approximate and exact values of $\lambda_n(p)$ start to match well for higher values of $n$. We note that the partial sum $\sum_{n=2}^{1000} \lambda_n(p)$ is equal to 0.978, 0.970, 0.955 and 0.910 for $p = 0.2, 0.4, 0.6$ and $0.8$, respectively; therefore, if the value of the erasure probability $(p)$ of the BEC is increased, then the tail of the sequence $\{\lambda_n(p)\}$ indeed becomes more significant (since the sum of all the $\lambda_n(p)$'s is 1). More specifically, we use the fact that

$$\sum_{n=N}^{\infty} n^{-\alpha} = \frac{N^{1-\alpha}}{\alpha-1}(1+o(1)), \quad \alpha > 1$$

and rely on Eq. (73) in order to show that for a large enough value of $N$, the sum $\sum_{n=2}^{N} \lambda_n(p)$ is approximately equal to $1 - \frac{1}{\sqrt{\pi N}}\frac{1}{1-p}$; this matches very well with the numerical values computed for $N = 1000$ with $p = 0.2$ and $0.8$.

# 6 Practical Considerations and Simulation Results

In this section, we present simulation results for both the bit-regular (Theorem 1) and check-regular (Theorem 2) ensembles. While these results are provided mainly to validate the claims of the theorems, we do compare them with one other previously known ensemble. This is meant to give the reader some sense of their relative performance. Note that for *fixed* complexity, the new codes eventually (for $n$ large enough) outperform any code proposed to date. On the other hand, the *convergence speed* to the ultimate performance limit is expected to be quite slow, so that for moderate lengths, the new codes are not necessarily expected to be record breaking.

## 6.1 Construction and Performance of Bit-Regular IRA Codes

The bit-regular plot in Fig. 4 compares systematic IRA codes [4] with $\lambda(x) = x^2$ and $\rho(x) = x^{36}$ (i.e., rate 0.925) with bit-regular non-systematic codes formed by our construction in Theorem 1 with $q = 3$. This comparison with non-systematic IRA codes was chosen for two reasons. First, both codes have good performance in the error floor region because neither have degree 2 information bit. Second, LDPC codes of such high rate have a large fraction of degree 2 bits and the resulting comparison seemed rather unfair. We remind the reader that the bit-regular ensembles of IRA codes in Theorem 1 are limited to high rates (for $q = 3$, the rate should be at least $\frac{12}{13} \approx 0.9231$).

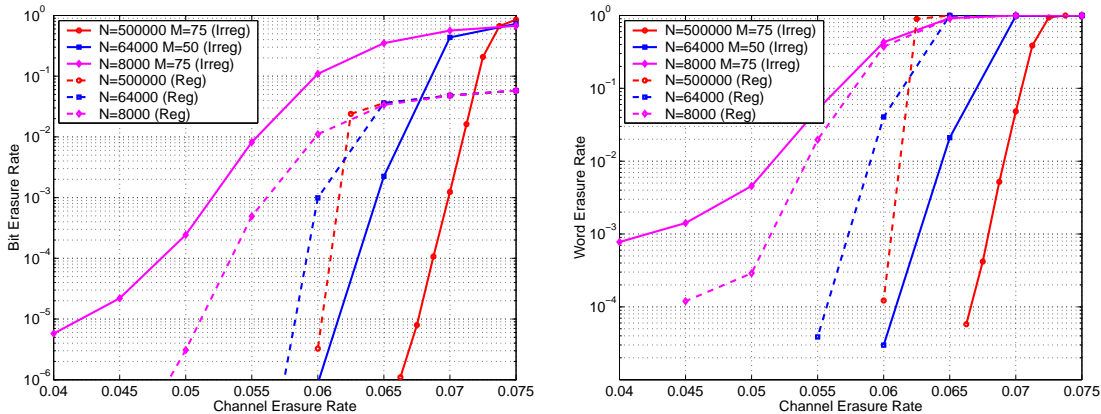

Figure 4: BER and WER for random rate 0.925 codes from the bit-regular IRA ensemble in Theorem 1 with $q = 3$ and the regular systematic IRA ensemble with d.d. $\lambda(x) = x^2$ and $\rho(x) = x^{36}$. The curves are shown for $N = 8000, 64000$, and $500000$.

The fixed point at $x = 1$ of the DE equation for the ensemble in Theorem 1 prevents the decoder from getting started without the help of some kind of "doping". We overcome this problem by using a small number of systematic bits (100–200) in the construction. Of course, these bits are included in the final rate of 0.925. Codes of block length $N = 8000, 64000$ and $500000$ were chosen from these ensembles and simulated on the BEC. The parity-check d.d. of each bit-regular code was also truncated to maximum degree $M$. All codes were constructed randomly from their d.d.s while avoiding 4 cycles in the subgraph induced by excluding the code bits (i.e., w.r.t. the top half of Fig. 1).

It is clear that the bit-regular ensemble performs quite well when the block length is large. As the block length is reduced, the fraction of bits required for "doping" (i.e., to get decoding started) increases and the performance is definitely degraded. In fact, the regular systematic IRA codes even outperform the bit-regular construction for a block length of 8000.

## 6.2 Construction and Performance of Check-Regular IRA Codes

The performance of the check-regular construction in Theorem 2 was also evaluated by simulation. A fixed rate of 1/2 was chosen and non-systematic IRA codes were generated with varying block length and maximum information-bit degree. For comparison, LDPC codes from the check-regular capacity-achieving ensemble [16] were constructed in the same manner. This ensemble was chosen for comparison because it has been shown to be essentially optimal for LDPC codes in terms of the tradeoff between performance and complexity [9, 16]. The IRA code ensembles were formed by treating all information bits degree greater than $M = 25, 50$ as pilot bits. The LDPC code ensembles were formed by choosing the check degree to be $q = 8, 9$ and then truncating the bit d.d. so that $\lambda(1) = 1$. This approach leads to maximum bit degrees of $M = 61, 126$, respectively. Actual codes of length $N = 8192, 65536$ and $524288$ were chosen from these ensembles, and simulated over the BEC. The results of the simulation are shown in Fig. 5. To simplify the presentation, only the best performing curve (in regards to truncation length $M$) is shown for each block length.

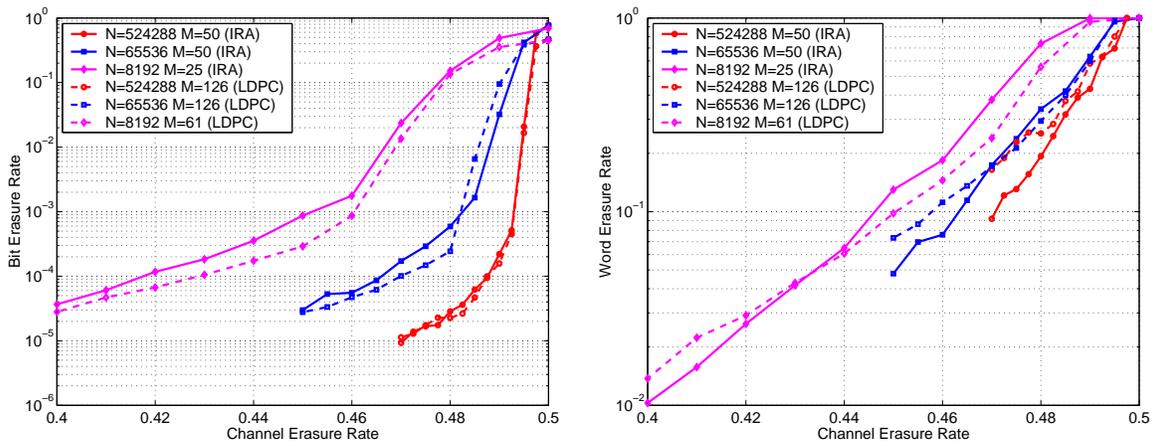

Figure 5: BER and WER for random rate 1/2 codes from the check-regular IRA ensemble in Theorem 2 and the check-regular LDPC ensemble [16] for $N = 8192, 65536$, and $524288$.

The code construction starts by quantizing the d.d. to integer values according to the block length. Next, it matches bit edges with check edges in a completely random fashion. Since this approach usually leads both multiple edges and 4-cycles, a post-processor is used. One iteration of post-processing randomly swaps all edges involved in a multiple-edge or 4-cycle events. We note that this algorithm only considers 4-cycles in the subgraph induced by excluding the code bits (i.e., the top half of the graph in Fig. 1). This iteration is repeated until there are no more multiple edges or 4-cycles. For the IRA codes, a single "dummy" bit is used to collect all of the edges originally destined for bits of degree greater than $M$. Since this bit is known to be zero, its column is removed to complete the construction. After this removal, the remaining IRA code is no longer check regular because this "dummy" bit is allowed to have multiple edges. In fact, it is exactly the check nodes which are reduced to degree 1 that allow decoding to get started. Finally, both the check-regular IRA and LDPC codes use an extended Hamming code to protect the information bits. This helps to minimize the effect of small weaknesses in the graph and improves the word erasure rate (WER) quite a bit for a very small cost. The rate loss associated with this is not considered in the stated rate of 1/2.

## 6.3 Stability Conditions

While the condition (7) is both necessary and sufficient for successful decoding, we can still gain some insight by studying (7) at its endpoints $x = 0$ and $x = 1$. The condition that the fixed point at $x = 0$ be stable is commonly known as the stability condition. Our capacity-achieving d.d. pairs actually satisfy (7), but focusing on the points $x = 0$ and $x = 1$ gives rise to just two stability conditions. For decoding to finish, the fixed point at $x = 0$ must be *stable*. While, to get decoding started, it helps if the fixed point at $x = 1$ is *unstable*.

The stability condition at $x = 0$ can be found by requiring that the derivative of the LHS of (7) is less than unity at $x = 0$. Writing this in terms of $\lambda_2$ (assuming that $\lambda(0) = 0$) gives

$$\lambda_2 < \frac{1}{\frac{2pR'(1)}{1-p} + \rho'(1)}. \tag{77}$$

Since capacity-achieving codes must satisfy the stability condition with equality, we see that the RHS of (77) must go to zero for capacity-achieving ensembles without degree 2 information bits. It is worth noting that the bit-regular ensemble in Theorem 1 (which has no degree 2 information bits) has $\rho'(1) = \infty$ and therefore meets the stability condition with equality.

The stability condition at $x = 1$ can be found by requiring that small deviations from $x = 1$ grow larger. Essentially, this means that the derivative of the LHS of (7) should be greater than 1 at $x = 1$. Writing this in terms of $\rho_2$ (assuming $\rho(0) = 0$) gives

$$\rho_2 > \frac{1}{(1-p)^2 \lambda'(1)}. \tag{78}$$

Since capacity-achieving codes must also satisfy this stability condition with equality, we see that the RHS of (78) must go to zero for capacity-achieving ensembles without degree 2 parity-check nodes. It is worth noting that the check-regular ensemble in Theorem 2 (which has no degree 2 parity-check nodes) has $\lambda'(1) = \infty$ and therefore meets the stability condition with equality. Furthermore, one can see intuitively how degree 2 parity-checks help to keep the decoding chain reaction from ending.

In reality, the fixed point at $x = 1$ is usually made unstable by either allowing degree 1 parity-checks or adding systematic bits (which has a similar effect). Therefore, the derivative condition is not strictly required. Still, it can play an important role. Consider what happens if you truncate the bit d.d. of the check-regular ensemble and then add systematic bits to get decoding started. The fixed point at $x = 1$ remains stable because the truncated bit d.d. has $\lambda'(1) < \infty$. In fact, the decoding curve in the neighborhood of $x = 1$ has a shape that requires a large number of systematic bits get decoding started reliably. This is the main reason that we introduced the "pilot bit" truncation in Theorem 2.

# 7   Conclusions

In this work, we present two sequences of ensembles of non-systematic irregular repeat-accumulate (IRA) codes which asymptotically (as their block length tends to infinity) achieve capacity on the binary erasure channel (BEC) with *bounded complexity* (throughout this paper, the complexity is normalized per information bit). These are the first capacity-achieving ensembles with bounded complexity on the BEC to be reported in the literature. All previously reported capacity-achieving sequences have a complexity which grows at least like the log of the inverse of the gap (in rate) to capacity. This includes capacity-achieving ensembles of LDPC codes [7, 8, 16], systematic IRA codes [4, 6, 15], and Raptor codes [17]. The ensembles of non-systematic IRA codes which are considered in our paper fall in the framework of multi-edge type LDPC codes [13].

We show that under message-passing iterative (MPI) decoding, this new bounded complexity result is only possible because we allow a sufficient number of state nodes in the Tanner graph representing a code ensemble. The state nodes in the Tanner graph of the examined IRA ensembles are introduced by puncturing all the information bits. We also derive an information-theoretic lower bound on the decoding complexity of randomly punctured codes on graphs. The bound refers to MPI decoding, and it is valid for an arbitrary memoryless binary-input output-symmetric channel with a special refinement for the BEC. Since this bound holds with probability 1 w.r.t. a randomly chosen puncturing pattern, it remains an interesting open problem to derive information-theoretic bounds that can be applied to *every puncturing pattern*. Under MPI decoding and the random puncturing assumption, it follows from the information-theoretic bound that a necessary condition to achieve the capacity of the BEC with bounded complexity or to achieve the capacity of a general memoryless binary-input output-symmetric channel with bounded complexity per iteration is that the puncturing rate of the information bits goes to one. This is consistent with the fact that the capacity-achieving IRA code ensembles introduced in this paper are non-systematic, where all the information bits of these codes are punctured.

In Section 6, we use simulation results to compare the performance of our ensembles to the check-regular LDPC ensembles introduced by Shokrollahi [16]. For the cases tested, the performance of our check-regular IRA codes is slightly worse than that of the check-regular LDPC codes. It is clear from these results that the fact that these capacity-achieving ensembles have bounded complexity does not imply that their performance, for small to moderate block lengths, is superior to other reported capacity-achieving ensembles. Note that for *fixed* complexity, the new codes eventually (for $n$ large enough) outperform any code proposed to date. On the other hand, the *convergence speed* to the ultimate performance limit happens to be quite slow, so for small to moderate block lengths, the new codes are not necessarily record breaking. Further research into the construction of codes with bounded complexity is likely to produce codes with better performance for small to moderate block lengths.

The central point in this paper is that by allowing state nodes in the Tanner graph, one may obtain a significantly better tradeoff between performance and complexity as the gap to capacity vanishes. Hence, it answers in the affirmative a fundamental question which was posed in [14] regarding the impact of state nodes (or in general, more complicated graphical models than bipartite graphs) on the performance versus complexity tradeoff under MPI decoding. Even the more complex graphical models, employed by systematic IRA codes, provides no asymptotic advantage over codes which are presented by bipartite graphs under MPI decoding (see [14, Theorems 1, 2] and [15, Theorems 1, 2]). Non-systematic IRA codes do provide, however, this advantage over systematic IRA codes; this is because the complexity of systematic IRA codes becomes unbounded, under MPI decoding, as the gap to capacity goes to zero.

# Appendices

## Appendix A: Proof of the Non-Negativity of the Power Series Expansion of $\rho(\cdot)$ in (10)

Based on the relation (1) between the functions $R(\cdot)$ and $\rho(\cdot)$, we see that $\rho(\cdot)$ has a non-negative power series expansion if and only if $R(\cdot)$ has the same property. We find it more convenient to prove that $R(\cdot)$ has a non-negative power series expansion. Starting with Eq. (28), we can rewrite $R(x)$ as

$$
\begin{aligned}
R(x) &= \frac{1}{p} \frac{\frac{p}{1-p} Q(x)}{1 + \frac{p}{1-p} Q(x)} \\
&= -\frac{1}{p} \sum_{i=1}^{\infty} \left( \frac{-p}{1-p} Q(x) \right)^i \\
&= -\frac{1}{p} \sum_{i=0}^{\infty} \left\{ \left( \frac{-p}{1-p} Q(x) \right)^{2i+1} + \left( \frac{-p}{1-p} Q(x) \right)^{2i+2} \right\} \\
&= \frac{1}{p} \left[ \frac{p}{1-p} Q(x) - \left( \frac{p}{1-p} Q(x) \right)^2 \right] \sum_{i=0}^{\infty} \left\{ \left( \frac{p}{1-p} Q(x) \right)^{2i} \right\}.
\end{aligned}
$$

One can verify from Eq. (62) that the power series coefficients of $Q(x)$ are positive. Therefore, the sum

$$
\sum_{i=0}^{\infty} \left\{ \left( \frac{p}{1-p} Q(x) \right)^{2i} \right\}
$$

also has a non-negative power series expansion. Based on this, it follows that the $R(\cdot)$ has a non-negative power series expansion if the function

$$
\frac{p}{1-p} Q(x) - \left( \frac{p}{1-p} Q(x) \right)^2
$$

has the same property. This means that the power series expansion of $R(\cdot)$ has non-negative coefficients as long as

$$
[x^k] \frac{p}{1-p} Q(x) \geq [x^k] \left( \frac{p}{1-p} Q(x) \right)^2 \quad k = 0, 1, 2, \ldots \tag{A.1}
$$

where $[x^k] A(x)$ is the coefficient of $x^k$ in the power series expansion of $A(x)$. Since $Q(\cdot)$ has a non-negative power series expansion starting from $x^2$, it follows that $Q^2(\cdot)$ has a non-negative power series expansion starting from $x^4$. Therefore, the condition in inequality (A.1) is automatically satisfied for $k < 4$. For $k = 4, 5$, the requirement in (A.1) leads (after some algebra) to the inequality in (18). Examining this condition for $q = 3$ and in the limit as $q \to \infty$, we find that the value of $p$ should not exceed $\frac{1}{13}$ and $\frac{3}{13}$, respectively. While we believe that the power series expansion of $R(\cdot)$ is indeed positive for all $p$ satisfying (18), we were unable to prove this analytically. Even if this is true, it follows that $R(\cdot)$ has a non-negative power series expansion only for rather small values of $p$.

For the particular case of $q = 3$, however, we show that the condition $p \leq \frac{1}{13}$ is indeed sufficient to ensure that (A.1) is satisfied for all $k \geq 0$.

*Proof.* In the case where $q = 3$, we find that

$$Q(x) = (-2 + 2x)\left(1 - \sqrt{1-x}\right) + x \tag{A.2}$$

and

$$Q(x)^2 = (8 - 20x + 12x^2)\left(1 - \sqrt{1-x}\right) - 4x + x^2 - 4x^3. \tag{A.3}$$

Expanding (A.2) in a power series gives

$$Q(x) = x + (-2 + 2x) \sum_{j=1}^{\infty} \left\{ \binom{\frac{1}{2}}{j} (-1)^{j+1} x^j \right\}$$

and matching terms shows that for $k \geq 2$

$$[x^k] Q(x) = \frac{6}{2k-3} \binom{\frac{1}{2}}{k} (-1)^{k+1}.$$

Doing the same thing for (A.3) shows that, for $k \geq 4$,

$$[x^k] Q(x)^2 = \left(8 - \frac{20k}{k - \frac{3}{2}} + \frac{12k(k-1)}{\left(k - \frac{3}{2}\right)\left(k - \frac{5}{2}\right)}\right) \binom{\frac{1}{2}}{k} (-1)^{k+1}.$$

The maximal value of $p$ such that the condition (A.1) is satisfied for $k \geq 0$ is given by

$$\frac{p}{1-p} \leq \frac{[x^k] Q(x)}{[x^k] Q(x)^2}$$

$$= \frac{\frac{6}{2k-3}}{8 - \frac{20k}{k-\frac{3}{2}} + \frac{12k(k-1)}{\left(k-\frac{3}{2}\right)\left(k-\frac{5}{2}\right)}}$$

$$= \frac{2k-5}{4(k+5)} \quad k = 4, 5, 6, \ldots$$

Since the RHS of this inequality is strictly increasing for $k \geq 4$, the maximal value of $p$ which satisfies (A.1) is found by substituting $k = 4$. This gives the condition $p \leq \frac{1}{13}$ and completes the proof that the power series expansion of $R(\cdot)$ is non-negative for $q = 3$ and $p \leq \frac{1}{13}$. □

## Appendix B: Proof of Properties of the D.D. Coefficients $\{\lambda_n\}$

The main part of this appendix proves the three different representation for the d.d. $\{\lambda_n\}$ which follows from the power series expansion of $\lambda(\cdot)$ in (15). We also consider in this appendix some properties of the polynomials $P_n(\cdot)$ which are related to the third representation of the d.d. $\{\lambda_n\}$, and which are also useful for the proof in Appendix C. Finally, we consider the asymptotic behavior of the d.d. $\{\lambda_n\}$.

We start this Appendix by proving the first expression for the d.d. $\{\lambda_n\}$ in (64)–(65).

*Proof.* The first representation of the d.d. $\{\lambda_n\}$ in (64) follows directly from the Lagrange inversion formula by simply writing $y\phi(x) = x$ where $y \triangleq \lambda^{-1}(x)$ is introduced in (39), and $\phi(\cdot) \triangleq \frac{x}{\lambda^{-1}(x)}$ is introduced in (65). We note that $\lambda_n$ is the coefficient of $x^{n-1}$ in the power series expansion of the function $\lambda(x)$ in (15). $\square$

The proof of the second expression for the d.d. $\{\lambda_n\}$ in (66)–(68) relies on the Cauchy residue theorem.

*Proof.* The function $\lambda(\cdot)$ in (15) is analytic except for three branch cuts shown in Fig. 6. The first branch cut starts at one and continues through the real axis towards infinity. The remaining two branches are straight lines which are symmetric w.r.t. the real axis. They are located along the lines defined by $z = 1 + c(p)re^{\pm \frac{i\pi}{3}}$, where $c(p) \triangleq \left(\frac{4(1-p)^3}{27p}\right)^{2/3}$ and $r \geq 1$. By the Cauchy Theorem we have

$$\lambda_n = \frac{1}{2\pi i} \oint_\Gamma \frac{\lambda(z)}{z^n} \, dz, \quad n \geq 2 \tag{B.1}$$

where the contour $\Gamma = \Gamma_1 \cup \Gamma_2 \cup \Gamma_3$ is the closed path which is shown in Fig. 6 and which is composed of three parts: the first part of integration ($\Gamma_1$) is parallel to the branch cut at an angle of 60°, and it starts from the point $z = 1$ on the real axis and goes to infinity along this line, so it can be parameterized as $\left\{\Gamma_1 : z = 1 + c(p)\, r\, e^{\frac{i\pi}{3}} + iw\right\}$ where $0 \leq r \leq R$ and $w \to 0^+$ is an arbitrarily small positive number (the straight line $\Gamma_1$ is parallel but slightly above the branch cut whose angle is 60° with the real axis). We later let $R$ tend to infinity. The second path of integration ($\Gamma_2$) is along the part of the circle of radius $\left|1 + c(p)Re^{\frac{i\pi}{3}}\right|$ where we integrate counter-clockwise, starting from $z = 1 + c(p)Re^{\frac{i\pi}{3}} + iw$ and ending at $z = 1 + c(p)Re^{-\frac{i\pi}{3}} - iw$. The third part of integration is a straight line which is parallel to the branch cut whose angle with the real axis is $-60°$, but is slightly below this branch cut; it starts at the point $z = 1 + c(p)Re^{-\frac{i\pi}{3}} - iw$, and ends at the point $z = 1$ on the real axis (so the points on $\Gamma_3$ are the complex conjugates of the points on $\Gamma_1$, and the directions of the integrations over $\Gamma_1$ and $\Gamma_3$ are opposite). Overall, $\Gamma \triangleq \Gamma_1 \cup \Gamma_2 \cup \Gamma_3$ forms a closed path of integration which does not contain the three branch cuts of the function $\lambda(\cdot)$ in (15), and it is analytic over the domain which is bounded by the contour $\Gamma$. We will first show that the above integral over $\Gamma_2$ vanishes as the radius of the circle tend to infinity, so in the limit where $R \to \infty$, we obtain that the integral over $\Gamma$ is equal to the sum of the two integrals over $\Gamma_1$ and $\Gamma_3$. In order to see that, we will show that the modulus of $\lambda(z) - 1$ in (15) is bounded over the circle $|z| = R$ (when $R$ is large), so it will yield that for $n \geq 2$, the integral $\int_{\Gamma_2} \frac{\lambda(z)-1}{z^n} \, \mathrm{d}z$ vanishes as $R \to \infty$. To this end, we rely on the equality

$$\sin\left(\frac{1}{3}\arcsin(z)\right) = \frac{1}{2i}\left[\left(iz + \sqrt{1-z^2}\right)^{\frac{1}{3}} - \left(iz - \sqrt{1-z^2}\right)^{-\frac{1}{3}}\right]$$

which yields that $\left|\sin\left(\frac{1}{3}\arcsin(z)\right)\right| = O\left(|z|^{\frac{1}{3}}\right)$.

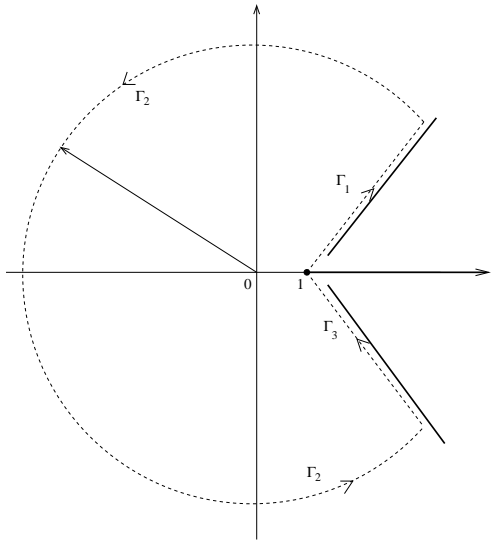

Figure 6: The branch cuts of the function $\lambda(x)$ in (15), and the contour of integration in Eq. (B.1).

By substituting $z' = \frac{3\sqrt{3}}{2}\sqrt{\frac{p(1-z)^{\frac{3}{2}}}{(-1+p)^3}}$, then

$$\left|\sin\left[\frac{1}{3}\arcsin\left(\frac{3\sqrt{3}}{2}\sqrt{\frac{p(1-z)^{\frac{3}{2}}}{(-1+p)^3}}\right)\right]\right| = O\left(|z'|^{\frac{1}{3}}\right) = O\left(|z|^{\frac{1}{4}}\right)$$

and from (15),

$$|\lambda(z) - 1| = \left|\sqrt{-\frac{4(1-p)}{3p\sqrt{1-z}}} \sin\left(\frac{1}{3}\arcsin\left(\sqrt{-\frac{27p(1-z)^{3/2}}{4(1-p)^3}}\right)\right)\right| = O\left(|z|^{-\frac{1}{4}}|z|^{\frac{1}{4}}\right) = O(1).$$

From the last equality, it follows that

$$\lim_{R \to \infty} \int_{\Gamma_2} \frac{\lambda(z)-1}{z^n} \, \mathrm{d}z = 0, \quad \forall\, n \geq 2. \tag{B.2}$$

Now we will evaluate the integral over the straight line $\Gamma_1$ (and similarly over $\Gamma_3$). Let

$$z = 1 + \left(\frac{4}{27}\frac{(1-p)^3}{p}\right)^{\frac{2}{3}} re^{\frac{i\pi}{3}} + iw, \quad w \to 0^+,\, r \geq 0$$

then after a little bit of algebra, one can verify that

$$\lambda(z) - 1 = \left(\frac{4}{p}\right)^{\frac{1}{3}} \cdot \frac{e^{\frac{2\pi i}{3}} \sin\left(\frac{1}{3}\arcsin(r^{\frac{3}{4}})\right)}{r^{\frac{1}{4}}} \tag{B.3}$$

so from the parameterization of $\Gamma_1$ and (B.3), we obtain that

$$\lim_{R \to \infty} \int_{\Gamma_1} \frac{\lambda(z)-1}{z^n} \, \mathrm{d}z = -\left(\frac{4}{p}\right)^{\frac{1}{3}} c(p) \int_0^{+\infty} \frac{g(r)}{\left(1 + c(p)\, r\, e^{\frac{i\pi}{3}}\right)^n} \, \mathrm{d}r \tag{B.4}$$

where $c(p)$ is introduced in (67), and the function $g(\cdot)$ is introduced in (68).

Since the points on $\Gamma_3$ are the complex conjugates of the points on $\Gamma_1$, and the integrations over $\Gamma_1$ and $\Gamma_3$ are in opposite directions, then it follows from (B.4) that

$$\lim_{R\to\infty} \int_{\Gamma_3} \frac{\lambda(z)-1}{z^n}\, dz = \left(\frac{4}{p}\right)^{\frac{1}{3}} c(p) \int_0^{+\infty} \frac{g^*(r)}{\left(1+c(p)\, r\, e^{-\frac{i\pi}{3}}\right)^n}\, dr. \tag{B.5}$$

By combining Eqs. (B.1)–(B.5) and (67) [we note that $c(p)$ in (67) is real for $0 < p < 1$], we obtain that for $n \geq 2$

$$\begin{aligned}
\lambda_n &= \frac{1}{2\pi i} \oint_\Gamma \frac{\lambda(z)}{z^n}\, dz \\
&= \frac{1}{2\pi i} \left( \lim_{R\to\infty} \int_{\Gamma_1} \frac{\lambda(z)-1}{z^n}\, dz + \lim_{R\to\infty} \int_{\Gamma_3} \frac{\lambda(z)-1}{z^n}\, dz \right) \\
&= \left(\frac{4}{p}\right)^{\frac{1}{3}} \frac{c(p)}{2\pi i} \left[ \int_0^{+\infty} \frac{g^*(r)}{\left(1+c(p)\, r\, e^{-\frac{i\pi}{3}}\right)^n}\, dr - \int_0^{+\infty} \frac{g(r)}{\left(1+c(p)\, r\, e^{\frac{i\pi}{3}}\right)^n}\, dr \right] \\
&= \left(\frac{4}{p}\right)^{\frac{1}{3}} \frac{c(p)}{2\pi i} \cdot (-2i)\, \mathrm{Im}\left\{ \int_0^{+\infty} \frac{g(r)}{\left(1+c(p)\, r\, e^{\frac{i\pi}{3}}\right)^n}\, dr \right\} \\
&= -\frac{c(p)}{\pi} \left(\frac{4}{p}\right)^{\frac{1}{3}} \mathrm{Im}\left\{ \int_0^{+\infty} \frac{g(r)}{\left(1+c(p)\, r\, e^{\frac{i\pi}{3}}\right)^n}\, dr \right\} \\
&= -\frac{4(1-p)^2}{9\pi p} \mathrm{Im}\left\{ \int_0^{+\infty} \frac{g(r)}{\left(1+c(p)\, r\, e^{\frac{i\pi}{3}}\right)^n}\, dr \right\}
\end{aligned}$$

which coincides with the representation of $\lambda_n$ in (66)–(68). $\square$

The Proof of the third expression for the sequence $\{\lambda_n\}$ in (69)–(75) is based on the previous expression which was proved above, and it enables to calculate the d.d. $\{\lambda_n\}$ in an efficient way.

*Proof.* From equation (66) which we already proved, then we obtain that

$$\begin{aligned}
\sum_{k=n+1}^\infty \lambda_k(p) &= -\frac{4(1-p)^2}{9\pi p} \mathrm{Im}\left\{ \int_0^{+\infty} g(r) \sum_{k=n+1}^\infty \frac{1}{\left(1+c(p)\, re^{\frac{i\pi}{3}}\right)^k}\, dr \right\} \\
&= -\frac{4(1-p)^2}{9\pi p} \mathrm{Im}\left\{ \int_0^{+\infty} \frac{g(r)}{1 - \frac{1}{1+c(p)\, re^{\frac{i\pi}{3}}}} \frac{1}{\left(1+c(p)\, r\, e^{\frac{i\pi}{3}}\right)^{n+1}}\, dr \right\} \\
&= -\frac{4(1-p)^2}{9\pi p\, c(p)} \cdot \mathrm{Im}\left\{ \int_0^{+\infty} \frac{g(r)}{r\, e^{\frac{i\pi}{3}} \left(1+c(p)\, r\, e^{\frac{i\pi}{3}}\right)^n}\, dr \right\} \\
&= -\left(\frac{4}{p}\right)^{\frac{1}{3}} \frac{1}{\pi} \cdot \mathrm{Im}\left\{ \int_0^{+\infty} \frac{g(r)}{r\, e^{\frac{i\pi}{3}} \left(1+c(p)\, r\, e^{\frac{i\pi}{3}}\right)^n}\, dr \right\}
\end{aligned}$$

where the last transition is based on (67). Therefore, by multiplying both sides of the last equality by $\left(\frac{p}{4}\right)^{\frac{1}{3}}$ and differentiating with respect to $p$, we obtain the equality

$$\frac{\partial}{\partial p}\left\{\left(\frac{p}{4}\right)^{\frac{1}{3}} \sum_{k=n+1}^{\infty} \lambda_k(p)\right\} = -\frac{1}{\pi} \cdot \mathrm{Im}\left\{\int_0^{+\infty} \frac{g(r)}{r\,e^{\frac{i\pi}{3}}} \cdot \frac{\partial}{\partial p}\left(\left(1+c(p)\,r\,e^{\frac{i\pi}{3}}\right)^{-n}\right)\partial r\right\}$$

$$= \frac{n}{\pi}\frac{\partial c}{\partial p} \cdot \mathrm{Im}\left\{\int_0^{+\infty} \frac{g(r)}{\left(1+c(p)\,r\,e^{\frac{i\pi}{3}}\right)^{n+1}}\partial r\right\}$$

$$= -\frac{n}{\pi}\frac{\partial c}{\partial p}\frac{9\pi p}{4(1-p)^2} \cdot \lambda_{n+1}(p)$$

where the last transition is based on (66). The function $c(\cdot)$ introduced in (67) is monotonic decreasing in $(0, 1]$, and

$$\frac{\partial c}{\partial p} = -\frac{2^{\frac{7}{3}}(1-p)(1+2p)}{27\,p^{\frac{5}{3}}} \qquad 0 < p < 1$$

so by combing the last equality with the previous one, it follows easily that

$$\frac{\partial}{\partial p}\left\{\left(\frac{p}{4}\right)^{\frac{1}{3}} \sum_{k=n+1}^{\infty} \lambda_k(p)\right\} = n\,\alpha(p)\,\lambda_{n+1}(p) \tag{B.6}$$

where

$$\alpha(p) \triangleq -\frac{9p}{4(1-p)^2}\frac{\partial c}{\partial p} = \frac{2^{\frac{1}{3}}}{3}\frac{1+2p}{1-p}\frac{1}{p^{\frac{2}{3}}}. \tag{B.7}$$

From (B.6), we get that for $n \geq 2$

$$\frac{\partial}{\partial p}\left\{\left(\frac{p}{4}\right)^{\frac{1}{3}}\lambda_n(p)\right\} = \frac{\partial}{\partial p}\left\{\left(\frac{p}{4}\right)^{\frac{1}{3}} \sum_{k=n}^{\infty}\lambda_k(p)\right\} - \frac{\partial}{\partial p}\left\{\left(\frac{p}{4}\right)^{\frac{1}{3}} \sum_{k=n+1}^{\infty}\lambda_k(p)\right\}$$

$$= \alpha(p)\left[(n-1)\,\lambda_n(p) - n\,\lambda_{n+1}(p)\right]$$

so, the following recursive equation follows for $n \geq 2$

$$\lambda_{n+1}(p) = \frac{1}{n}\left[(n-1)\lambda_n(p) - \frac{1}{\alpha(p)}\frac{\partial}{\partial p}\left\{\left(\frac{p}{4}\right)^{\frac{1}{3}}\lambda_n(p)\right\}\right]$$

$$\stackrel{(a)}{=} \frac{1}{n}\left(n-1-\frac{1}{2}\frac{1-p}{1+2p}\right)\cdot\lambda_n(p) - \frac{3}{2n}\frac{p(1-p)}{1+2p}\cdot\lambda'_n(p)$$

$$= \frac{1}{2(1+2p)n}\left[\left((4n-3)p+(2n-3)\right)\lambda_n(p) - 3p(1-p)\lambda'_n(p)\right]$$

where transition (a) is based on (B.7), and the last two transitions involve a little bit of algebra. In order to start the recursive equation from $n = 1$, we write it as

$$\lambda_{n+2}(p) = \frac{[(4n+1)p+(2n-1)]\,\lambda_{n+1}(p) - 3p(1-p)\lambda'_{n+1}(p)}{2(1+2p)(n+1)} \qquad n = 1, 2, \ldots \tag{B.8}$$

with the initial value $\lambda_2(p) = \frac{1-p}{2(1+2p)}$. Based on the recursive equation (B.8) and the value of $\lambda_2(\cdot)$, it can be shown that

$$\lambda_3(p) = \frac{(1-p)(1+16p+10p^2)}{8(1+2p)^3}$$

$$\lambda_4(p) = \frac{(1-p)(1+12p+168p^2+164p^3+60p^4)}{16(1+2p)^5}$$

and so on, which suggests the possible substitution $\lambda_{n+1}(p) = \frac{(1-p)P_n(p)}{(1+2p)^{2n-1}}$ for a certain sequence of polynomials $\{P_n(\cdot)\}_{n\geq 1}$. Combining the last substitution with (B.8) gives rather easily the recursive equation (72), and it therefore justifies the existence of such polynomials $P_n(\cdot)$ in (71) which can be calculated recursively from (72) with the initial polynomial $P_1(\cdot) = \frac{1}{2}$ (the initial polynomial is determined from the value of $\lambda_2(\cdot)$). In general, it is easy to verify from (72) that for $n \geq 1$, $P_n(\cdot)$ is a polynomial of degree $2(n-1)$. Eq. (75) follows easily from (72), as if we substitute $p = 0$ or $p = 1$ in (72), then the coefficient of $P_n'(\cdot)$ vanishes, so we obtain that

$$P_{n+1}(0) = \frac{2n-1}{2(n+1)} P_n(0), \qquad P_{n+1}(1) = \frac{9(2n+1)}{2(n+1)} P_n(1), \qquad n = 1, 2, \ldots$$

where $P_1(0) = P_1(1) = \frac{1}{2}$. The closed form solutions for $P_n(0)$ and $P_n(1)$ are given in (75). □

We will prove now a property of the polynomials $P_n(\cdot)$ which will be useful for the proof of the positivity of the power series expansion of $\lambda(\cdot)$ (see Appendix C).

**Lemma 3.** If for some $n \in \mathbb{N}$, the polynomial $P_n(\cdot)$ has a zero in the interval $[0,1]$, then the polynomial $P_{n+1}(\cdot)$ has also a zero in the same interval.

*Proof.* Since from Eq. (75), $P_n(0)$ and $P_n(1)$ are both positive, then if $P_n(\cdot)$ has a zero in $[0,1]$, then it follows that the minimal value of $P_n(\cdot)$ over this interval is obtained at an interior point $x_n \in (0,1)$. So, $P_n(x_n) \leq 0$ and $P_n'(x_n) = 0$. Since $(14 - 4n)x^2 + (20n - 4)x + (2n - 1) > 0$ for $n \geq 1$ and $x \in [0,1]$, then from Eq. (72), $P_{n+1}(x_n) \leq 0$. From Eq. (75), $P_{n+1}(0)$ and $P_{n+1}(1)$ are both positive, but $P_{n+1}(x_n) \leq 0$, so the continuity of the polynomial $P_{n+1}(\cdot)$ yields that this polynomial has at least one zero inside the interval $[0,1]$. As a direct consequence of this proof and the equality in (71), we obtain that if for a certain positive integer $n$, the value of $\lambda_n(\cdot)$ is positive in the interval $[0,1)$, then for $k = 2, 3, \ldots, n-1$, $\lambda_k(\cdot)$ should be also positive in this interval. □

The above observation points towards one viable way of proving the positivity of the sequence $\{\lambda_n\}$: show that $P_n(\cdot)$ is strictly positive for all $x \in [0,1]$ for $n$ sufficiently large. The following heuristic argument shows that this should indeed be the case. Unfortunately though, it seems nontrivial to make this argument precise, and therefore we will use in Appendix C a different route to actually prove the positivity of the coefficients.

Our heuristic argument goes as follows. Consider Fig. 7 (see next page) which shows $\frac{\ln(P_n(\cdot))}{n}$ for increasing values of $n$. This figure suggests that $\frac{\ln(P_n(x))}{n}$ converges uniformly over the whole range $x \in [0,1]$ to a smooth limit. In order to find this limiting function, let us assume that approximately

$$P_n(x) \doteq C_n e^{nf(x)} \quad \text{where} \quad \lim_{n \to \infty} \frac{\ln(C_n)}{n} = 0.$$

Then we obtain that

$$\begin{aligned} P_{n+1}(x) &\doteq C_{n+1} e^{(n+1)f(x)} = C_{n+1} e^{nf(x)} e^{f(x)} \\ P_n'(x) &\doteq C_n n e^{nf(x)} f'(x). \end{aligned}$$

By substituting these asymptotic expressions in (72), we obtain the equation

$$\left(\frac{C_{n+1}}{C_n}\right) e^{f(x)} = \frac{(14 - 4n)x^2 + (20n - 4)x + (2n - 1) - 3nx(1 + x - 2x^2)f'(x)}{2(n+1)}. \tag{B.9}$$

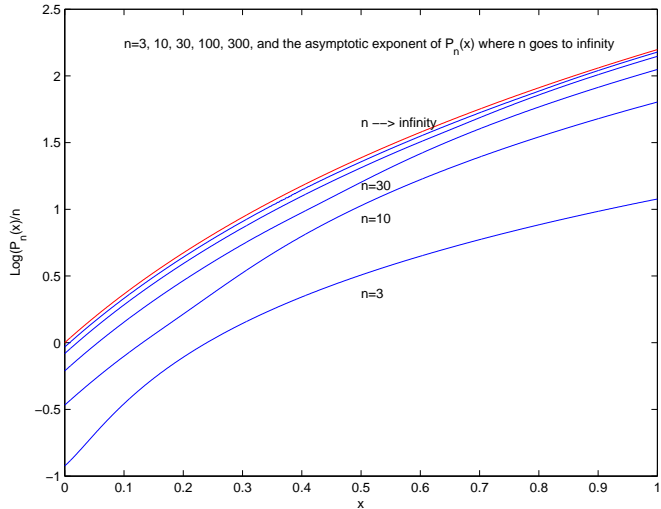

Figure 7: Plots of $\frac{\ln(P_n(x))}{n}$ for the polynomials $P_n(\cdot)$ which are introduced in Eq. (72). In the limit where $n \to \infty$, these curves converge uniformly on $[0,1]$ to the function $f(x) = 2\ln(1+2x)$.

Let us assume that $\lim_{n\to\infty} \frac{C_{n+1}}{C_n} = 1$ (this assumption will be verified later). Then in the limit where $n \to \infty$, (B.9) yields the following differential equation

$$e^{f(x)} = -2x^2 + 10x + 1 - \frac{3}{2} x(1 + x - 2x^2) f'(x). \tag{B.10}$$

From (B.10), it follows directly that $f(0) = 0$ and $f(1) = \ln(9)$. It can be easily verified that the solution of the differential equation (B.10) is $f(x) = 2\ln(1+2x)$ which in spite of the approximation suggested above coincides with the limiting function of $\frac{\ln(P_n(x))}{n}$ when $n$ tends to infinity (see Fig. 7). We note that the assumption that $\frac{C_{n+1}}{C_n}$ converges to 1 can be justified from (75) as follows: for sufficiently large $n$

$$C_n \approx \frac{P_n(1)}{e^{nf(1)}} = \frac{P_n(1)}{9^n} = \frac{1}{9} \frac{1}{4^n} \binom{2n}{n} = O\left(\frac{1}{\sqrt{n}}\right)$$

so the limit of $\frac{C_{n+1}}{C_n}$ is equal to 1 as $n$ tends to infinity. In a neighborhood of $x = 1$, we obtain from the solution for the asymptotic function $f(\cdot)$ that

$$P_n(x) \doteq \frac{1}{9} \frac{1}{4^n} \binom{2n}{n} (1+2x)^{2n}. \tag{B.11}$$

We note that after after some straightforward (though tedious) calculations, it can be shown from (72) that $P_n'(1) = \frac{12 \cdot 9^{n-2} (n-1)}{4^n} \binom{2n}{n}$, and hence it follows from (75) that $P_n'(1) = \frac{4(n-1)}{3} P_n(1)$. However, the approximation in (B.11) gives that $P_n'(1) \approx \frac{4n}{3} P_n(1)$ which is indeed a very good approximation to the equality that we get from an accurate analysis (especially for large values of $n$). We also note that if we fix the value of $n$ (for a large enough value of $n$), and we let $p$ tend to unity (i.e., $p \to 1^-$), then it follows from (72) and (B.11) that

$$\lambda_{n+1}(p) = \frac{(1-p)P_n(p)}{(1+2p)^{2n-1}}$$

$$\approx \frac{1}{3} \frac{1}{4^n} \binom{2n}{n} (1-p)$$

which coincides with (76).

*The asymptotic behavior of the d.d.* $\{\lambda_n\}$: We will discuss now the asymptotic behavior of the d.d. $\{\lambda_n\}$ which is given in Section 5.2.2. We refer the reader to [3] which relates an asymptotic expansion of a function around a dominant singularity with the corresponding asymptotic expansion for the Taylor coefficients of the function. The problem in our case is that in order to get good approximations of the power series expansion of $\lambda(\cdot)$ in (15) around the dominant singularity (which is at 1), the function can be extended analytically beyond the radius of 1 for as far as possible except for a cone around 1 (see the solid lines in Fig. 6 which are the branch cuts of $\lambda(\cdot)$). The further one can expand the function around the dominant singularity, the more it is determined by this singularity, and the Taylor series expansion around this singularity will be very accurate for determining the behavior of $\lambda_n$ starting from moderate values of $n$. For the function $\lambda(\cdot)$ in (15), if we increase the value of $p$ and make it closer to 1, then unfortunately, the two other singularities of $\lambda(\cdot)$ at $z = 1 + c(p)e^{\pm \frac{i\pi}{3}}$, where $c(\cdot)$ is given in (67), move towards 1 very quickly (e.g., if $p = 0.8$, then the other two singularities are located at $1.0065 \pm 0.011255i$). This is the reason why by increasing the value of $p$, the asymptotic expansion kicks in for larger values of $n$ (see Fig. 3).

The proof of the asymptotic behavior of the d.d. $\{\lambda_n\}$ goes as follows. If $z \to 1$, then the function $\lambda(\cdot)$ in (15) is approximately equal to $1 - \frac{\sqrt{1-z}}{1-p}$ (since if $u \approx 0$, then $\sin(u) \approx u$ and $\arcsin(u) \approx u$). The coefficient of $z^n$ in the latter function is $\frac{(-1)^{n+1}}{1-p} \binom{\frac{1}{2}}{n}$ which is equal to $\frac{1}{1-p} \frac{1}{2n-1} \frac{1}{4^n} \binom{2n}{n}$. Therefore, the contribution of the dominant singularity of $\lambda(\cdot)$ to the asymptotic behavior of $\lambda_{n+1}$ (i.e., the coefficient of $z^n$ in the Taylor series expansion of $\lambda(z)$) is given by

$$\frac{1}{1-p} \frac{1}{2n-1} \frac{1}{4^n} \binom{2n}{n}$$

$$= \frac{n^{-\frac{3}{2}}}{2\sqrt{\pi}(1-p)} \left(1 + \frac{3}{8}\frac{1}{n} + \frac{25}{128}\frac{1}{n^2} + O\left(\frac{1}{n^3}\right)\right). \tag{B.12}$$

This already gives the first term of the asymptotic behavior of $\lambda_n$ in (73); based on our explanation above, it is a tight approximation of the asymptotic behavior of $\lambda_{n+1}(p)$ for rather small values of $p$. As was mentioned above, for larger values of $p$, the other two singularities at $z_{1,2} = 1 + c(p)e^{\pm \frac{i\pi}{3}}$ where $c(\cdot)$ is introduced in (67) are very close to the dominant singularity at $z = 1$. In order to determine the asymptotic behavior of $\lambda_{n+1}(p)$ for these larger values of $p$, we therefore need to take into account the asymptotic expansion of the function $\lambda(\cdot)$ around these two singularities. After some algebra, one can verify that the behavior of the function $\lambda(\cdot)$ around its singularity at $z = z_1$ is like

$$\lambda(z) \approx A(z) \triangleq 1 + \frac{e^{i\frac{2\pi}{3}}}{\sqrt[3]{2p}} - \frac{3i}{2\sqrt{2}} \frac{\sqrt{z_1}\sqrt{1 - \frac{z}{z_1}}}{1-p},$$

and the behavior of $\lambda(\cdot)$ around its second singularity at $z = z_2$ is like the complex conjugate of $A(z)$ (since $z_1$ and $z_2$ form a pair of complex conjugates). The influence of these two singularities on the asymptotic behavior of $\lambda_{n+1}(p)$ is therefore equal to $2\,\mathrm{Re}\{[z^n]A(z)\}$, where $[z^n]A(z)$ designates the coefficient of $z^n$ in the power series expansion of $A(z)$. Calculation shows that for $n \geq 1$

$$2\,\mathrm{Re}\{[z^n]A(z)\} = \frac{3\sqrt{2}}{2(1-p)} \frac{1}{2n-1} \frac{1}{4^n} \binom{2n}{n} a_p^{-(n-\frac{1}{2})} \sin\left(\left(n - \frac{1}{2}\right)\theta_p\right)$$

where $a_p$ and $\theta_p$ are introduced in (74), and for large values of $n$, the last expression behaves like

$$\frac{3}{4}\sqrt{\frac{2}{\pi}} \frac{n^{-\frac{3}{2}}}{1-p} a_p^{-(n-\frac{1}{2})} \sin\left(\left(n-\frac{1}{2}\right)\theta_p\right)\left(1 + \frac{3}{8}\frac{1}{n} + \frac{25}{128}\frac{1}{n^2} + O\left(\frac{1}{n^3}\right)\right). \tag{B.13}$$

The sum of the two terms in (B.12) and (B.13) finally gives the asymptotic behavior of $\lambda_{n+1}(p)$ in (73).

## Appendix C: Proof of the Positivity of the Power Series Expansion of $\lambda(\cdot)$ in (15)

The function $\lambda(\cdot)$ in (15) is analytic except for three branch cuts shown in Fig. 8. The first branch cut starts at one and continues towards infinity. The remaining two branches are symmetric around the real axis. They are located along the line $x = 1 + c(p)re^{\pm\frac{i\pi}{3}}$ where $c(p) = \left(\frac{4(1-p)^3}{27p}\right)^{2/3}$ and $r \geq 1$. By the Cauchy Theorem we have

$$\lambda_{n+1} = \frac{1}{2\pi i} \oint \frac{\lambda(z)}{z^{n+1}} dz,$$

where the contour can be taken e.g., along a circle of radius $r$, $r < 1$, enclosing the origin. Such a circle is shown as dashed line in Fig. 8. Also shown is a modified contour of integration in which the circle is expanded so that in the limit the integral wraps around the three branch cuts (we note that the path of integration in Fig. 8 is clearly different from that one in Fig. 6). Note that the modulus of $\lambda(z)$ is bounded so that for large values of $R$, the integral around the circle vanishes as $R \to \infty$ [in Appendix B, we show that if $R \to \infty$ then $|\lambda(z)| \leq O(1)$ on the circle $|z| = R$, so $\left|\frac{\lambda(z)}{z^{n+1}}\right| \leq O\left(\frac{1}{R^{n+1}}\right)$.]

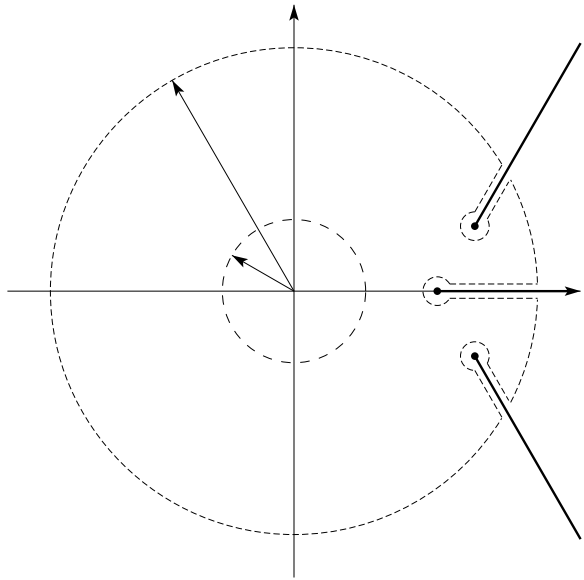

Figure 8: The branch cuts of the function $\lambda(x)$ in (15), the original contour of integration of radius $r$, $r < 1$, and the modified contour with radius $R$ tending to infinity.

Taking advantage of the symmetries in the problem, we see after a little bit of calculus that

$$\lambda_{n+1} = \text{Re}\left\{\frac{-8(1-p)^2}{9p\pi} \int_1^\infty \frac{h(r)}{\left(1+c(p)re^{\frac{\pi i}{3}}\right)^{n+1}} dr + \lim_{\epsilon \to 0^+} \frac{1}{\pi i} \int_1^\infty \frac{\lambda(x+\epsilon i)}{x^{n+1}} dx\right\}, \quad \text{(C.1)}$$

where $c(p)$ is introduced in (67) and

$$h(r) \triangleq \lim_{\alpha \to 0^+} \text{Im}\left\{\frac{\sin\left(\frac{1}{3}\arcsin(r^{\frac{3}{4}}e^{i\alpha})\right)}{r^{\frac{1}{4}}}\right\}, \quad r \geq 1. \quad \text{(C.2)}$$

Since

$$\lim_{\epsilon \to 0^+} \text{Re} \left\{ \frac{1}{\pi i} \int_1^\infty \frac{\lambda(x + \epsilon i)}{x^{n+1}} \, \mathrm{d}x \right\}$$

$$= \lim_{\epsilon \to 0^+} \text{Im} \left\{ \frac{1}{\pi} \int_1^\infty \frac{\lambda(x + \epsilon i)}{x^{n+1}} \, \mathrm{d}x \right\}$$

$$= \lim_{\epsilon \to 0^+} \text{Im} \left\{ \frac{1}{\pi} \int_1^\infty \frac{\lambda(x + \epsilon i) - 1}{x^{n+1}} \, \mathrm{d}x \right\}$$

$$= \lim_{\epsilon \to 0^+} \frac{1}{\pi} \int_1^\infty \frac{\text{Im}\{\lambda(x + \epsilon i) - 1\}}{x^{n+1}} \, \mathrm{d}x$$

then for $n \geq 1$, we can rewrite (C.1) in the equivalent form

$$\lambda_{n+1} = \text{Re} \left\{ \frac{-8(1-p)^2}{9p\pi} \int_1^\infty \frac{h(r)}{\left(1 + c(p) r e^{\frac{\pi i}{3}}\right)^{n+1}} \mathrm{d}r \right\} + \lim_{\epsilon \to 0^+} \frac{1}{\pi} \int_1^\infty \frac{\text{Im}\{\lambda(x + \epsilon i) - 1\}}{x^{n+1}} \, \mathrm{d}x \quad \text{(C.3)}$$

and the function $h(\cdot)$ in (C.2) can be expressed in a simpler way as

$$h(r) = \frac{\sqrt{3}}{4} \left[ \left(1 + \sqrt{1 - r^{-\frac{3}{2}}}\right)^{\frac{1}{3}} - \left(1 - \sqrt{1 - r^{-\frac{3}{2}}}\right)^{\frac{1}{3}} \right], \; r \geq 1. \quad \text{(C.4)}$$

Although we will not make use of this in the sequel, we note that in this representation, the function smoothly interpolates $\lambda_n$ for non-integral values of $n$.

We start by bounding the absolute value of the first term in the RHS of (C.3). From (C.4), it follows immediately that $h(r)$ is positive and monotonic increasing in $r$ for $r \geq 1$, and that it is upper bounded by $2^{-\frac{5}{3}}\sqrt{3}$ (which is the limit of $h(r)$ as $r \to \infty$). We therefore get from the non-negativity of $c(p)$ in (67) that

$$\left| \text{Re} \left\{ \frac{-8(1-p)^2}{9p\pi} \int_1^\infty \frac{h(r)}{\left(1 + c(p) r e^{\frac{\pi i}{3}}\right)^{n+1}} \mathrm{d}r \right\} \right|$$

$$\leq \frac{2^{\frac{4}{3}}(1-p)^2}{3\sqrt{3} p\pi} \int_1^\infty \left| 1 + c(p) r e^{\frac{\pi i}{3}} \right|^{-(n+1)} \mathrm{d}r$$

$$= \frac{2^{\frac{4}{3}}(1-p)^2}{3\sqrt{3} p\pi} \int_1^\infty \left(1 + c(p)^2 r^2 + c(p) r\right)^{-\frac{n+1}{2}} \mathrm{d}r$$

$$\leq \frac{2^{\frac{4}{3}}(1-p)^2}{3\sqrt{3} p\pi} \int_1^\infty \left(1 + c(p) r\right)^{-\frac{n+1}{2}} \mathrm{d}r$$

$$= \frac{2\sqrt{3}\left(1 + c(p)\right)^{-\frac{n-1}{2}}}{p^{\frac{1}{3}} \pi (n-1)}. \quad \text{(C.5)}$$

We note that (C.5) decreases by at most a factor $\frac{1}{\sqrt{1+c(p)}}$ when we increase $n$ to $n+1$. We use this result later in (C.12).

Regarding the second integral in the RHS of (C.3), we note that for $x \geq 1$

$$\lim_{\epsilon \to 0^+} \frac{1}{\pi} \operatorname{Im}\{\lambda(x+\epsilon i) - 1\} = \frac{2}{\pi}\sqrt{\frac{1-p}{3p}} \; \frac{t\left((x-1)^{\frac{1}{4}} \frac{3\sqrt{3p}}{2(1-p)^{\frac{3}{2}}}\right)}{(x-1)^{\frac{1}{4}}} \qquad \text{(C.6)}$$

where

$$t(u) \triangleq \operatorname{Im}\left\{ e^{\frac{i3\pi}{4}} \sin\left(\frac{1}{3}\arcsin\left(ue^{-\frac{i\pi}{4}}\right)\right) \right\}, \quad u \geq 0. \qquad \text{(C.7)}$$

Now, we rely on the inequality (which is shown in Fig. 9)

$$t(u) \geq f(u) \triangleq \frac{1}{3}u - \frac{16}{729}u^5, \quad u \geq 0. \qquad \text{(C.8)}$$

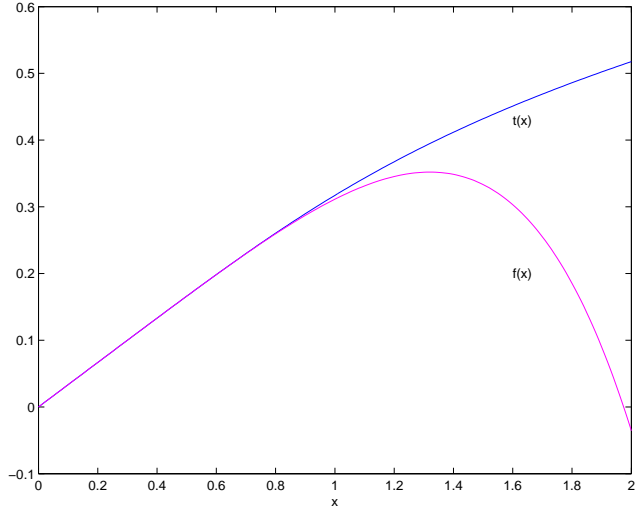

Figure 9: A plot which verifies graphically the inequality in (C.8).

It follows from (C.6), (C.7) and (C.8) that

$$\lim_{\epsilon \to 0^+} \frac{1}{\pi} \int_1^\infty \frac{\operatorname{Im}\{\lambda(x+\epsilon i) - 1\}}{x^{n+1}} dx$$

$$\geq \frac{2}{\pi}\sqrt{\frac{1-p}{3p}} \int_1^\infty \frac{f\left((x-1)^{\frac{3}{4}} \frac{3\sqrt{3p}}{2(1-p)^{\frac{3}{2}}}\right)}{(x-1)^{\frac{1}{4}} x^{n+1}} dx$$

$$= \frac{1}{\pi(1-p)} \int_1^\infty \frac{(x-1)^{\frac{1}{2}}}{x^{n+1}} dx - \frac{3p^2}{\pi(1-p)^7} \int_1^\infty \frac{(x-1)^{\frac{7}{2}}}{x^{n+1}} dx$$

$$= \frac{\Gamma(n-\frac{1}{2})}{2(1-p)\sqrt{\pi}\,\Gamma(n+1)} - \frac{315 p^2\, \Gamma(n-\frac{7}{2})}{16(1-p)^7 \sqrt{\pi}\, \Gamma(n+1)}. \qquad \text{(C.9)}$$

Equality (C.9) follows from the identity

$$\int_1^\infty \frac{(x-1)^\alpha}{x^{n+1}} dx = \frac{\Gamma(\alpha+1)\,\Gamma(n-\alpha)}{\Gamma(n+1)}, \quad -1 < \alpha < n, \qquad \text{(C.10)}$$

where $\Gamma(\cdot)$ designates the complete Gamma function [the integral in the LHS of (C.10) can be calculated by the substitution $x = \frac{1}{1-u}$ which transforms it to $\int_0^1 (1-u)^{n-\alpha-1} u^\alpha \, du$, and then the latter integral is by definition equal to $B(\alpha+1, n-\alpha)$ where $B(\cdot, \cdot)$ designates the Beta function; this function is related to the complete Gamma function by the equality $B(m,n) = \frac{\Gamma(m)\Gamma(n)}{\Gamma(m+n)}$.]. Since $\Gamma(x+1) = x\Gamma(x)$, and $\Gamma\left(\frac{1}{2}\right) = \sqrt{\pi}$, then we obtain from (C.10) that

$$\int_1^\infty \frac{(x-1)^{\frac{1}{2}}}{x^{n+1}} \, dx = \frac{\Gamma\left(\frac{3}{2}\right) \Gamma\left(n - \frac{1}{2}\right)}{\Gamma(n+1)} = \frac{\sqrt{\pi} \, \Gamma\left(n - \frac{1}{2}\right)}{2 \, \Gamma(n+1)}$$

$$\int_1^\infty \frac{(x-1)^{\frac{7}{2}}}{x^{n+1}} \, dx = \frac{\Gamma\left(\frac{9}{2}\right) \Gamma\left(n - \frac{7}{2}\right)}{\Gamma(n+1)} = \frac{\left(\frac{7}{2} \cdot \frac{5}{2} \cdot \frac{3}{2} \cdot \frac{1}{2} \cdot \sqrt{\pi}\right) \Gamma\left(n - \frac{1}{2}\right)}{\Gamma(n+1)} = \frac{105}{16} \frac{\sqrt{\pi} \, \Gamma\left(n - \frac{1}{2}\right)}{\Gamma(n+1)}$$

which confirms the equality in (C.9).

We note that (C.9) (which forms a lower bound on the second term in the RHS of (C.3)) decreases by at most a factor $1 - \frac{3}{2(n+1)}$ uniformly over $p$ when we increase $n$ to $n+1$ (it can be shown by using the recursive equation $\Gamma(x+1) = x\Gamma(x)$ for the complete Gamma function).

From (C.3), (C.5) and (C.9) and since $c(\cdot)$ in (67) is a monotonic decreasing and non-negative function over the interval $(0,1)$, it now follows that if for a specific $n^* \in \mathbb{N}$ and for all $p \in [0, p^*]$

$$\frac{\Gamma(n^* - \frac{1}{2})}{2(1-p)\sqrt{\pi} \, \Gamma(n^* + 1)} - \frac{315 p^2 \, \Gamma(n^* - \frac{7}{2})}{16(1-p)^7 \sqrt{\pi} \, \Gamma(n^* + 1)} \geq \frac{2\sqrt{3}(1 + c(p))^{-\frac{n^* - 1}{2}}}{p^{\frac{1}{3}} \pi (n^* - 1)} \quad \text{(C.11)}$$

$$1 - \frac{3}{2(n^* + 1)} \geq \frac{1}{\sqrt{1 + c(p^*)}} \quad \text{(C.12)}$$

then $\lambda_n(p) > 0$ for all $n > n^*$ and $p \in [0, p^*]$.

In the following, we consider the positivity of the sequence $\{\lambda_n(p)\}$ for $p \in [0, 0.95]$. From the above bounds, it follows that we can prove the positivity of this sequence in some band $p \in [0, p^*]$ by checking the positivity of only a finite number of terms in this sequence (we note that this number grows dramatically for values of $p^*$ which are very close to 1). Assume we pick $p^* = 0.95$. By explicitly evaluating (C.11), we see that we need $n^* \geq 7957$. From condition (C.12), we get $n^* \geq 4144$, so a valid choice for the fulfillment of both conditions is $n^* = 7957$. Based on (C.11) and (C.12), we conclude that $\lambda_n(p) > 0$ for all $n > n^*$ and $p \in [0, p^*]$. For $n \leq n^*$ and all $p \in [0, 1)$, we will verify the positivity of the coefficients $\{\lambda_n(p)\}$ by alternatively showing that for $n \leq n^*$, the polynomials $P_n(\cdot)$ in (72) are positive in the interval $[0, 1]$. This equivalence follows from (71). First we can observe from (75) that since $P_n(0)$ and $P_n(1)$ are positive for all $n \in \mathbb{N}$, then the polynomials $P_n(\cdot)$ are positive in the interval $[0, 1]$ if and only if they have no zeros inside this interval. Hence, based on Lemma 3 (see Appendix B), one can verify the positivity of $P_n(p)$ for $n \leq n^*$ and $p \in [0, 1]$ by simply verifying the positivity of $P_{n^*}(\cdot)$ in the interval $[0, 1]$. To complete our proof, we proceed as follows. We write

$$P_n(p) = \sum_{i=0}^{2(n-1)} b_i^{(n)} (p - p_0)^i \quad \text{(C.13)}$$

where for convenience we choose $p_0 = \frac{1}{2}$ (this will be readily clarified). Based on (69) where $P_n(p) = \sum_{i=0}^{2(n-1)} a_i^{(n)} p^i$, it follows that

$$b_i^{(n)} = \frac{P_n^{(i)}(p_0)}{i!} = \sum_{j \geq i} \binom{j}{i} a_j^{(n)} p_0^{j-i} \, . \quad \text{(C.14)}$$

Therefore, from the recursive equation (70) for $\{a_i^{(n)}\}$ and from (C.14), it follows that all coefficients $\{b_i^{(n)}\}$ are rational and can be calculated from the coefficients $\{a_i^{(n)}\}$ which are defined recursively in (70) and are rational as well. Using an infinite precision package, those coefficients can be computed exactly. By explicit computation, we verify that all the coefficients $b_i^{(n)}$ are strictly positive for $n = n^*$ and $0 \leq i \leq 2(n^* - 1)$, and therefore it follows from (C.13) that $P_{n^*}(\cdot)$ is positive (and strictly increasing) in the interval $[p_0, 1]$. For $p \in [0, p_0]$, one can verify from the conditions in (C.11) and (C.12) that $\lambda_{n+1}(p)$ and $P_n(p)$ are positive for $n \geq 57$ and $p \in [0, p_0]$. Combining these results, we conclude that $P_{n^*}(\cdot)$ is positive in the interval $[0, 1]$. This therefore concludes the proof that $\lambda_n(p)$ is positive for all $n \in \mathbb{N}$ and $p \in [0, 0.95]$.

Though not proving the positivity of $\lambda_n(\cdot)$ over the whole interval $[0, 1)$, we note that the uniform convergence of the plots which are depicted in Fig. 7 (see Appendix B) and (71) strongly supports our conjecture about the positivity of $\lambda_n(\cdot)$ over this interval.

# Figure Captions

**Figure 1**: The Tanner graph of IRA codes.

**Figure 2**: The function $\lambda(\cdot)$ in (15), as a function of the erasure probability ($p$) of a BEC.

**Figure 3**: The exact and approximated values of $\lambda_{n+1}$ ($n = 1, \ldots, 1000$) for the check-regular IRA ensemble in Theorem 2. The approximation is based on the asymptotic expression in (73). The plots refer to a BEC whose erasure probability is $p = 0.2$ (left plot) and $p = 0.8$ (right plot).

**Figure 4**: BER and WER for random rate 0.925 codes from the bit-regular IRA ensemble in Theorem 1 with $q = 3$ and the regular systematic IRA ensemble with d.d. $\lambda(x) = x^2$ and $\rho(x) = x^{36}$. The curves are shown for $N = 8000, 64000$, and $500000$.

**Figure 5**: BER and WER for random rate $\frac{1}{2}$ codes from the check-regular IRA ensemble in Theorem 2 and the check-regular LDPC ensemble [16] for $N = 8192, 65536$, and $524288$.

**Figure 6**: The branch cuts of the function $\lambda(x)$ in (15), and the contour of integration in Eq. (B.1).

**Figure 7**: Plots of $\frac{\ln(P_n(x))}{n}$ for the polynomials $P_n(\cdot)$ which are introduced in Eq. (72). In the limit where $n \to \infty$, these curves converge uniformly on $[0, 1]$ to the function $f(x) = 2\ln(1 + 2x)$.

**Figure 8**: The branch cuts of the function $\lambda(x)$ in (15), the original contour of integration of radius $r$, $r < 1$, and the modified contour with radius $R$ tending to infinity.

**Figure 9**: A plot which verifies graphically the inequality in (C.8).